\documentclass[twocolumn]{aastex631}

\usepackage{amsmath}
\usepackage{makecell} 
\usepackage{enumitem}
\usepackage{ulem}

\begin{document}
\setlist[enumerate]{itemsep=0pt, topsep=0pt, parsep=0pt, partopsep=0pt}
\setlist[itemize]{itemsep=0pt, topsep=0pt, parsep=0pt, partopsep=0pt}

\title{Photometric Determination of Unresolved Main-sequence Binaries in the Pleiades: Binary Fraction and Mass Ratio Distribution}

\author[0000-0001-5968-1144]{Rongrong Liu}
\affiliation{Shanghai Astronomical Observatory, Chinese Academy of Sciences, \\
80 Nandan Road, Shanghai 200030, People's Republic of China}
\affiliation{School of Astronomy and Space Sciences, University of Chinese Academy of Sciences,\\ 
No. 19A Yuquan Road, Beijing 100049, People's Republic of China}

\author[0000-0001-8611-2465]{Zhengyi Shao}
\affiliation{Shanghai Astronomical Observatory, Chinese Academy of Sciences, \\
80 Nandan Road, Shanghai 200030, People's Republic of China}
\affiliation{Key Lab for Astrophysics, Shanghai 200234, People's Republic of China}

\author[0000-0002-0880-3380]{Lu Li}
\affiliation{Shanghai Astronomical Observatory, Chinese Academy of Sciences, \\
80 Nandan Road, Shanghai 200030, People's Republic of China}

\begin{abstract}
Accurate determination of binary fractions ($f_{\rm b}$) and mass ratio ($q$) distributions is crucial for understanding the dynamical evolution of open clusters. We present an improved multiband fitting technique to enhance the analysis of binary properties. This approach enables an accurate photometric determination of $f_{\rm b}$ and $q$ distribution in a cluster. The detectable mass ratio can be down to the $q_{\rm lim}$, limited by the minimum stellar mass in theoretical models. First, we derived an empirical model for magnitudes of Gaia DR3 and 2MASS bands that match the photometry of single stars in the Pleiades. We then performed a multiband fitting for each cluster member, deriving the probability density function (PDF) of its primary mass ($\mathcal{M}_1$) and $q$ in the Bayesian framework. 1154 main-sequence (MS) single stars or unresolved MS+MS binaries are identified as members of the Pleiades. By stacking their PDFs, we conducted a detailed analysis of binary properties of the cluster. We found the $f_{\rm b}$ of this sample is $0.34 \pm 0.02$. The $q$ distribution exhibits a three-segment power-law profile: an initial increase, followed by a decrease, and then another increase. This distribution can be interpreted as a fiducial power-law profile with an exponent of -1.0 that is determined in the range of $0.3 < q < 0.8$, but with a deficiency of binaries at lower $q$ and an excess at higher $q$. The variations of $f_{\rm b}$ and $q$ with $\mathcal{M}_1$ reveal a complex binary distribution within the Pleiades, which might be attributed to a combination of primordial binary formation mechanisms, dynamical interactions, and the observational limit of photometric binaries imposed by $q_{\rm lim} (\mathcal{M}_1)$.

\end{abstract}


\section{Introduction} \label{sec:intro}

The member stars of an open cluster (OC) formed in the same molecular cloud and thus constitute a single stellar population (SSP) \citep{2003ARA&A..41...57L}. Young open clusters typically contain a high proportion of primordial binary \citep{2023ASPC..534..275O}. With frequent three-body encounters, these binaries may dissolve or have their companions exchanged with other single stars, leading to changes in the statistical properties of binaries, such as the binary fraction ($f_{\rm b}$) and the mass ratio ($q$) distribution \citep{1975MNRAS.173..729H,2011MNRAS.417.1684M,2013ApJ...779...30G,2013AJ....145....8G,2015ApJ...805...11G}. Therefore, these binary properties, as important as the stellar mass function, provide critical constraints on the models of star formation and dynamical evolution in OCs.

The Pleiades cluster, one of the nearest OCs, contains over a thousand main-sequence stars, making it an ideal target for detecting and analyzing binaries. As early as eighty years ago, \cite{1944ApJ...100..360S} identified two Pleiades members as spectroscopic binaries. Over the past forty years, the $f_{\rm b}$ of the Pleiades has been discussed by numerous studies. But significantly divergent results have been yielded, from 13$\%$ to 73$\%$, depending on different stellar mass ranges and spatial regions \citep{1982AJ.....87.1507S,1992A&A...265..513M,1995MNRAS.272..630S,1997A&A...323..139B,2003ApJ...594..525M,2003MNRAS.342.1241P,2007MNRAS.380..712L,2020ApJ...901...91T,2023MNRAS.525.2315A,2023AJ....165...45M}.

Determining the mass ratio of a binary is another challenge. Traditionally, $q$ is defined as the dynamical mass ratio derived from radial velocity variations using spectroscopic observations. Unfortunately, this method is expensive and usually requires decades to gather sufficient high-precision data. For instance, despite 39 years of data accumulation, \cite{2021ApJ...921..117T} identified 48 binaries within a spectroscopic sample of 289 Pleiades members, of which only 21 had their dynamical mass ratios determined. Moreover, this method is less effective for binaries with lower mass ratios, longer orbital periods, or higher orbital inclinations, which will result in detection biases.

In contrast, using multiband photometric data to investigate those unresolved binaries is more efficient due to the extensive dataset and the convenience of establishing a complete magnitude-limited sample. For cluster members, since their distances, ages, metallicities, and dust extinctions are identical to those of the cluster, the magnitudes of a cluster member in all bands are entirely dependent on its mass, or the masses of its binary components. 

Several recent studies have investigated binaries in the Pleiades using  Gaia's photometric data on the color-magnitude diagram (CMD) \citep{2020ApJ...903...93N, 2021AJ....162..264J, 2023AA...672A..29C,  2023AA...675A..89D}. Despite the high-precision photometric data of Gaia, detecting binaries with $q<0.5$ remains difficult with optical data alone. \cite{2022AJ....163..113M} and \cite{2023AJ....165...45M} improved the detection by combining photometric data from optical to mid-infrared bands on a pseudo-color diagram, which can identify binaries including low mass-ratio ones. However, their studies were limited to stars within a mass range from $0.5\mathcal{M}_{\odot}$ to $1.8\mathcal{M}_{\odot}$. This limitation is caused by the potential bias in measurements of $q$, due to discrepancies between observational magnitudes and theoretical models at lower masses, as well as the difficulty of distinguishing binaries from single stars at higher masses.

The most straightforward method of employing multiband photometric data is taking a multiband magnitude fitting, which is broadly equivalent to the spectral energy distribution (SED) fitting. Some studies have used this technique to detect binaries in OCs by combining optical and infrared photometry. For example, \cite{2021AJ....161..160T} employed BINOCS to fit optical to mid-infrared magnitudes for stars in eight OCs. They accurately identified binaries with $q\gtrsim0.3$ within these clusters after correcting the model isochrone of SSP. Similarly, \cite{2024ApJ...962...41C} applied BASE-9, which is also a multiband fitting procedure, with optical to near-infrared (NIR) data to derive the binary fraction and mass-ratio distribution for six OCs. Their results were notably accurate for binaries with $q\gtrsim0.4$.

Regardless of the technique employed, detecting of low mass-ratio binaries remains extremely challenging. Nevertheless, this problem is very important for two reasons. First, it is essential for a comprehensive understanding of stellar mass distribution. Second, low mass-ratio binaries are crucial for studying dynamical effects due to their smaller binding energies, which make them more sensitive to dynamical interactions.

To address this problem, two key issues should be considered in photometric methods. One is the systematic deviation between theoretical model magnitudes and observational data, which can introduce biases in the measurement of binary mass ratios. Such deviations may be due to either the inaccuracies in models or the imperfect calibrations of observations. Despite substantial improvements in both theoretical models and observational techniques in recent years \citep{2014MNRAS.444.2525C,2014MNRAS.445.4287T,2015MNRAS.452.1068C,2016IAUFM..29B.144F,2018MNRAS.476..496F,2019A&A...632A.105C,2018A&A...615A..24W,2018A&A...616A...4E,2021A&A...649A...3R,2023A&A...674A...3M}, minor discrepancies remain and may affect on mass-ratio estimates. In studies of OCs, a common method to reduce these biases is to construct the so-called empirical isochrone. For example, one can measure the color differences between the observational main sequence and the model predictions on the CMD, and correct the model colors \citep{2012A&A...540A..16M,2019A&A...622A.110F,2020ApJ...901...49L}. If multiple bands are included, discrepancies of multiple colors will be measured. Then, the magnitudes of each band can be adjusted to a reference band, followed by multiband fitting \citep{2021AJ....161..160T}. An alternate approach is introducing error floors for individual observational magnitudes in the fitting. These additional errors should be large enough to cover discrepancies between models and observations, as well as the zero-point differences between photometric surveys \citep{2024ApJ...962...41C}.

The second issue arises from the lower limit of stellar mass ($\sim 0.075\mathcal{M}_{\odot}$). Companions with masses below this threshold, such as brown dwarfs or giant planets, do not contribute a detectable increment in luminosity. This factor leads to a lower limit on the mass ratio of photometric binaries, which depends on the mass of the primary star. Therefore, when discussing photometric binaries, it is crucial to specify the ranges of stellar masses and mass ratios. However, observational errors often make it difficult to distinguish low mass-ratio binaries from single stars, resulting in a higher detection limit of $q$. For instance, analyses based only on optical photometry usually adopt mass ratio limits between 0.5 and 0.7 \citep{2021AJ....162..264J,2023AA...672A..29C,2023AA...675A..89D}. Combining optical and infrared data can reduce this limit to $q\sim$ 0.2 to 0.4 \citep{2021AJ....161..160T,2022AJ....163..113M,2023AJ....165...45M,2024ApJ...962...41C}. Nonetheless, some low mass-ratio binaries remain challenging to be identified definitively.

In this work, we aim to apply novel strategies to address these two issues and investigate the stellar masses of Pleiades members, including the establishment of empirical photometric models, multiband fitting, and a study of unresolved main-sequence binaries in detail.

The rest of the paper is organized as follows. In Section~\ref{sec:data}, we introduce the observational data of the Pleiades, including the selection of kinematic members, the Gaia and 2MASS photometric data, and the fundamental physical parameters of the cluster. In Section~\ref{sec:method}, we first introduce the method for correcting discrepancies between observational magnitudes and the PARSEC model, presenting the empirical model of each adopted band for the Pleiades. Then, we describe the method and the result of multiband fitting and present the binary probability for each cluster member. In Section~\ref{sec:discussion}, we first describe the method for analyzing $f_{\rm b}$ and $q$ distribution by stacking probability density functions (PDFs) of all cluster members. Then, we discuss $f_{\rm b}$ and $q$ distribution in the Pleiades and their correlations with the primary mass. Finally, Section~\ref{sec:conclusion} summarizes this paper.

\section{Observational Data and Photometric Model of the Pleiades Cluster}\label{sec:data}
\begin{table*}
    \caption{Observational Data and Kinematic Membership Probabilities of Stars in the Region of the Pleiades.}
    \vspace{-10pt}
    \label{tab:data}
    \begin{center}
    \setlength{\tabcolsep}{8.7pt}
    \begin{tabular}{ccccccccc}
\hline
\hline
SourceId & RA & DEC & $\mu_{\rm{\alpha}}$ & $\mu_{\rm{\delta}}$ &  $\varpi$ & RUWE & $P_{\rm{k}}$ & $DM$\\
& (deg) & (deg) & (mas/yr) & (mas/yr)  & (mas)  & & & (mag)\\
(1) & (2) & (3) & (4) & (5)  & (6)  & (7) & (8) & (9)\\
\hline
68051390279853824  & 53.002 & 23.775 & 21.20 ± 0.02 & -43.69 ± 0.02  & 7.28 ±  0.02 & 1.069 & 0.993 & 5.689 \\
70461554129560448  & 57.052 & 26.584 & 20.98 ± 0.03 & -46.56 ± 0.02  & 7.63 ±  0.02 & 1.029 & 0.997 & 5.589 \\
71377000637253504  & 54.539 & 27.586 & 18.97 ± 0.09 & -43.57 ± 0.07  & 6.92 ±  0.08 & 5.014 & 0.798 & 5.769 \\
68587161680620288  & 55.496 & 25.329 & 20.10 ± 0.04 & -44.18 ± 0.03  & 7.38 ±  0.03 & 2.026 & 1.000 & 5.660 \\
64401389632616960  & 55.348 & 22.032 & 19.50 ± 0.09 & -44.14 ± 0.07  & 7.35 ±  0.07 & 0.983 & 0.999 & 5.665 \\
\hline
\end{tabular}

    \\[-3pt]
    \setlength{\tabcolsep}{15pt}
    \begin{tabular}{cccccc}
\hline
\hline
$G_{\rm{BP}}$ & $G$ & $G_{\rm{RP}}$ & $J$ & $H$ & $K_{\rm{s}}$ \\
(mag) & (mag)& (mag) & (mag) & (mag) & (mag)\\
(10) & (11) & (12) & (13) & (14)  & (15)\\
\hline
14.099 ± 0.005 & 13.306 ± 0.003 & 12.435 ± 0.005 & 11.33 ± 0.02 & 10.69 ± 0.02 & 10.52 ± 0.02  \\
13.949 ± 0.005 & 13.179 ± 0.003 & 12.313 ± 0.005 & 11.15 ± 0.02 & 10.49 ± 0.02 & 10.36 ± 0.02  \\
14.154 ± 0.003 & 13.341 ± 0.003 & 12.430 ± 0.004 & 11.24 ± 0.02 & 10.63 ± 0.02 & 10.47 ± 0.02  \\
14.373 ± 0.005 & 13.344 ± 0.003 & 12.326 ± 0.005 & 10.98 ± 0.02 & 10.30 ± 0.03 & 10.09 ± 0.02  \\
18.864 ± 0.032 & 16.889 ± 0.003 & 15.564 ± 0.005 & 13.56 ± 0.02 & 12.86 ± 0.03 & 12.54 ± 0.03  \\
\hline
\end{tabular}

    \end{center}
    \tablecomments{The first six columns list the Gaia DR3 source ID and astrometric data. The seventh column presents the RUWE which is commonly larger for binaries than single stars. The eighth column presents the kinematic membership probability, derived from the Gaia astrometric data. The ninth column shows the modified distance modulus (see Appendix~\ref{apen:BayesPlx}). Columns (10)-(15) provide photometric data from Gaia DR3 and 2MASS. (The complete version of this table is available in a machine-readable format.)}
\end{table*}
\begin{figure*}
    \includegraphics[width=\textwidth]{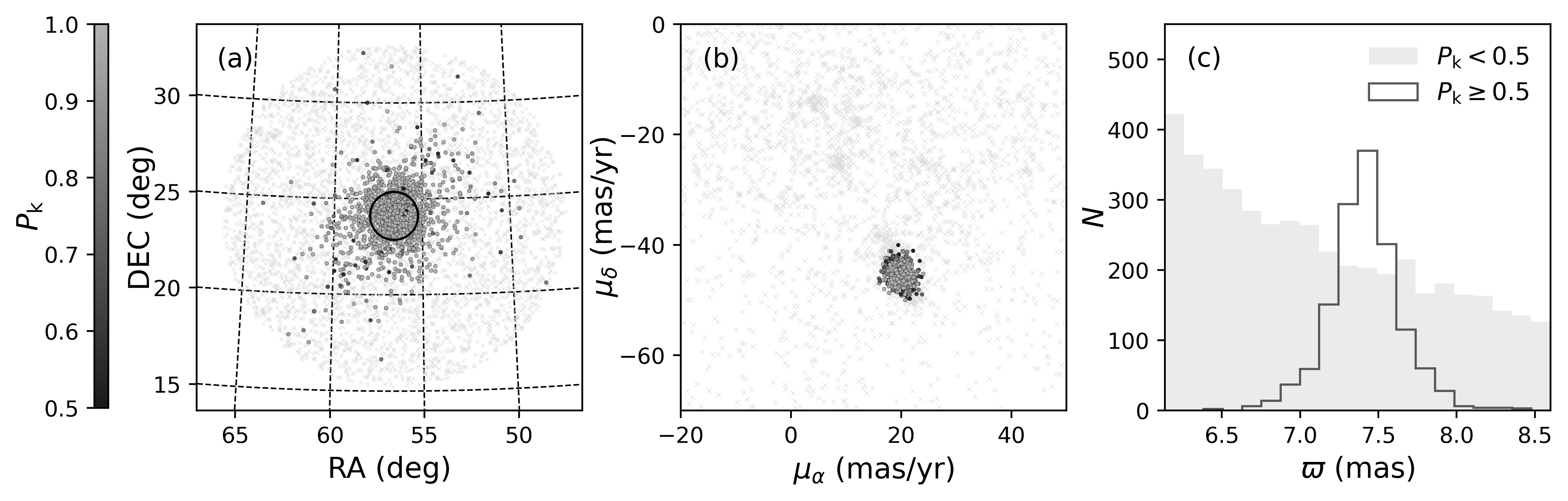}
    \caption{Distributions of equatorial coordinates (a), proper motions (b), and parallaxes (c) for stars in the Pleiades region. The crosses denote field stars with kinematic membership probability $P_{\rm k}<0.5$. The gray points represent kinematic member stars with $P_{\rm k}\geq0.5$. The black circle in panel (a) shows the half-number radius $R_{50}=1.37^\circ$ of the kinematic members.}
    \label{fig:data}
\end{figure*}

This section presents the observational data and photometric model used in this work, including kinematic members (Section~\ref{sec:member}), photometric data (Section~\ref{sec:photometry}), the fundamental physical parameters of the Pleiades and the corresponding fiducial photometric model (Section~\ref{sec:model}). 

\subsection{Sample and Kinematic Members}\label{sec:member} 

The Pleiades cluster is centered on $\alpha=56.601^\circ$, $\delta=24.114^\circ$, with a distance ranging from 130 to 140 pc \citep{2005A&A...429..645S,2014Sci...345.1029M} and a half-number radius of $\sim 1.37^\circ$ (approximately 3pc). Firstly, sample stars were restricted to $G<19$ mag and located within a radius of $8.9^\circ$ from the Pleiades center. For this flux-limited subset, $\sim 98.9\%$ of Gaia DR3 sources possess a 5-parameter astrometric solution. We further restricted the sample to parallaxes between 6.1 and 8.6 mas. This resulted in an initial sample of 6041 stars. Notably, this sample has not been constrained by the renormalized unit weight error (RUWE) value, because numerous unresolved binaries are among stars with RUWE$>$1.4. 

Subsequently, within the five-dimensional astrometric data space (equatorial coordinates, proper motions, and parallax) from Gaia DR3, we employed a two-component mixture model to determine the overall kinematic parameters of the cluster, and simultaneously calculate the kinematic membership probability ($P_{\rm k}$) for each sample star (Shao et al. 2024, in preparation). In this process, the observational errors of proper motions and parallax are taken into account in order to obtain the intrinsic dispersions in the space of these observables and the more rigorous statistical estimation of membership probabilities for individual stars. The astrometric data and $P_{\rm k}$ values of sample stars are listed in Table~\ref{tab:data}. Thanks to Gaia’s high-precision proper motions and parallaxes, a clear distinction between cluster members and field stars can be shown in the Pleiades region.  There are 97.6$\%$ sample stars having $P_{\rm k}>0.9$ or $P_{\rm k}<0.1$. Only a small fraction of objects remain ambiguous. The effectiveness index~\footnote{The index $E=1-N\cdot \sum_{i}{P_i(1-P_i)}/[\sum_i{P_i}\cdot\sum_i{(1-P_i)}]$, where $i=1,...,N$ for individual stars. $E=0$ indicates that the two components are totally mixed and cannot be separated anymore, whereas $E=1$ implies that the components are fully distinguishable \citep{1996AcASn..37..377S}. } of this kinematic membership determination is $E_{\rm k}=0.97$, which is very close to a complete separation of members and field stars.

A total of 1390 stars having $P_{\rm{k}}\geq0.5$  were chosen for subsequent analysis, with an average $P_{\rm k}$ value of 0.978. These members exhibit a high concentration in both coordinate and proper motion spaces, as shown in panels (a) and (b) of Figure~\ref{fig:data} respectively. Panel (c) demonstrates that the parallaxes of the members also follow a clustered distribution, distinctly different from that of field stars, which is in agreement with our expectations for a cluster region (see Equations (2) and (3) in \cite{2019MNRAS.489.3093S}).

Table~\ref{tab:member} compares our membership sample with several previous works also based on the Gaia astrometry \citep{2019AA...628A..66L,2020AA...633A..99C,2023AA...673A.114H,2024AJ....168..156C}. The comparison is limited to the same ranges of G-band magnitude, coordination, and parallax. Overall, our sample overlaps well with the previous samples. \cite{2019AA...628A..66L} and \cite{2023AA...673A.114H} have obtained significantly more members than ours; however, all of these unmatched members exhibit large proper motion dispersions and have low membership probabilities in their clustering algorithms.


Usually, for distant clusters, the distance of a member star can be approximated by the value of the cluster, as the distance variations among members are negligible. However, since the Pleiades is remarkably close to us, its member stars exhibit noticeable distance variations, corresponding to about 0.05 mag in distance modulus ($DM$), or about 3 pc in distance. Therefore, for the subsequent photometric fitting, we applied individual $DM$ for each star, derived from the modified Gaia parallax as described in Appendix~\ref{apen:BayesPlx}. The modified $DM$ values are listed in column (9) of Table~\ref{tab:data}.

\begin{table*}
    \caption{Comparison of Pleiades Members with Other Works.}
    \vspace{-10pt}
    \label{tab:member}
    \begin{center}
    \setlength{\tabcolsep}{11.5pt}
    \begin{tabular}{rrrccccc}
\hline
\hline
$N_{\rm TW}$  & $N_{\rm OW}$ & $N_{\rm{match}}$ & $G$ & Radius & Parallax & Source & Reference \\
&  &  & (mag) & (deg) & (mas) &  & \\
\hline
365  & 389 & 362 & $<15$ & $<2$ & 6.1 - 8.6 & Gaia DR3 & \cite{2024AJ....168..156C} \\
1123 & 1061 & 1027 & $<18$ & $<4$ & 6.9 - 7.9 & Gaia DR2 & \cite{2020AA...633A..99C} \\
1390 & 1616 & 1313 & $<19$ & $<8.9$ & 6.1 - 8.6 & Gaia DR2 & \cite{2019AA...628A..66L} \\
1390 & 1563 & 1353 & $<19$ & $<8.9$ & 6.1 - 8.6 & Gaia DR3 & \cite{2023AA...673A.114H}\\
\hline
\end{tabular}

    \end{center}
\end{table*}
\subsection{Photometric Data}\label{sec:photometry}

The photometric data used in this paper include $G_{\rm{BP}}$, $G$, $G_{\rm{RP}}$ from Gaia DR3~\citep{2016A&A...595A...1G,2023A&A...674A...1G,2023A&A...674A..32B}, and $J$, $H$, $K_{\rm s}$ from 2MASS~\citep{2006AJ....131.1163S}. This extensive coverage from optical to NIR bands improves the detection of low mass-ratio binaries compared to using optical bands alone. For Gaia BP and RP bands, the spatial resolution is 3.5 by 2.1 arcsec \citep{2023A&A...674A...2D}, while for 2MASS bands, the effective resolution is approximately 5 arcsec~\footnote{https://irsa.ipac.caltech.edu/data/2MASS/docs/releases/\\allsky/doc/sec2$\_$2a.html}. Therefore, the photometry of individual objects might be confused by their close neighbors. In order to eliminate this influence, we reject 153 cluster members having nearby Gaia sources within 5 arcsec radius. Of remaining 1237 kinematic members, 1231 have magnitudes for all six bands, representing 99.5$\%$ of the total. Six of the remainder have the Gaia photometric data alone. Detailed photometric data for each star are also provided in Table~\ref{tab:data}.

Typical photometric errors are about 0.003$\sim$0.005 mag for Gaia and 0.02$\sim$0.03 mag for 2MASS. However, the photometric uncertainties of Gaia are reported to be approximately 0.01 mag \citep{2021A&A...649A...3R}, which is significantly larger than the typical errors from the catalog. Using unreasonably high-precision photometry for some individual bands can lead to local convergence or bias in the multiband fitting. Therefore, we set an error floor of 0.01 mag for Gaia magnitudes. Thus, for each band of Gaia, the photometric error of each star is adjusted to $err=(err_{\rm obs}^2+0.01^2)^{1/2}$. This adjustment has an additional benefit of reducing the large difference between Gaia and 2MASS errors, thereby relatively increasing the weight of 2MASS data in the fitting process, making it more balanced with the optical photometry.

\begin{figure*}
    \includegraphics[width=\textwidth]{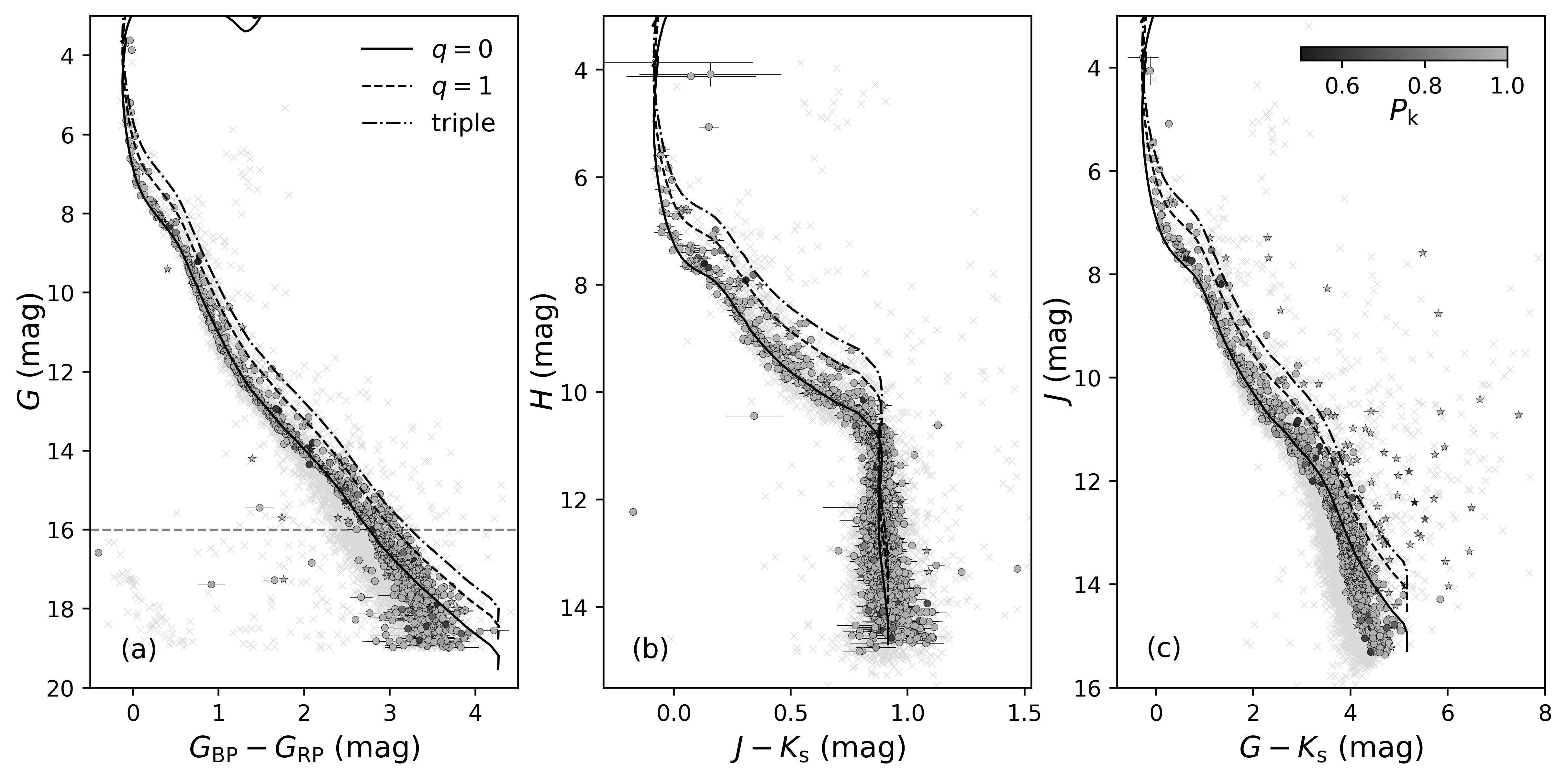}
    \caption{Distributions of stars in the Pleiades region on CMDs of Gaia (a), 2MASS (b), and combined optical and NIR bands (c). The crosses denote stars with kinematic membership probability $P_{\rm{k}}<0.5$. The gray points show stars with $P_{\rm{k}}\geq0.5$. The star symbols mark the stars with neighbors in 5 arcsec. The error bars represent original observational errors. The black solid and dashed lines represent isochrones for $q = 0$ and 1, respectively, from the PARSEC 1.2s model of [Fe/H]=0.1, $\log(Age)/yr=8.026$ and $A_{\rm V}=0.135$ mag. The dot-dash lines represent isochrons for triples with equal masses. The horizontal line marks $G=16$ mag. Note that three stars have been excluded from the plot due to invalid errors.}
    \label{fig:data_cmd}
\end{figure*}

Figure~\ref{fig:data_cmd} illustrates the distribution of sample stars on CMDs using Gaia, 2MASS, and one kind of combination of them. These CMDs clearly reveal the main sequence of single stars and the binary distribution of cluster members. Faint members exhibit greater scatter due to their larger errors. Some objects between the isochrones of binary and triple stars may be the multiple stellar systems in the Pleiades, which is not the focus of this paper. Outlier members to the lower left of the main sequence in panel (a) may represent member binaries consisting of a main-sequence star and a white dwarf (MS+WD), which will be investigated in a separate study due to their unique research interest. Most outliers appearing as discrete dots in panel (c) are those affected by the blending of their neighboring sources in the 2MASS bands, which leads to redder $G-K_{\rm s}$ colors. Others may be contaminated field stars. The different distribution patterns across these three CMDs demonstrate the necessity of combining optical and NIR photometry to identify non-cluster or particular cluster members, even though the kinematic membership method is highly effective. These stars, including field stars, MS+WD binaries and multiple stellar systems, will be subsequently excluded by the multiband fitting quality (see Section~\ref{sec:result} for details). In this paper, we focus only on MS+MS binaries of the Pleiades.

\subsection{Fundamental Parameters of the Pleiades and Its Fiducial theoretical Model}\label{sec:model}

In order to construct the empirical photometric model for the Pleiades, we need a well-defined fiducial theoretical model depending on the fundamental physical parameters of this cluster. In this work, we adopt the PARSEC1.2s model \citep{2012MNRAS.427..127B} due to its extensive range of age and metallicity. The model's minimum stellar mass\footnote{The stellar mass used through in this paper is the initial mass, whereas the corresponding present mass can be inferred according to the cluster's age.} is 0.09$\mathcal{M}_{\odot}$, enabling it to detect stars down to such a low mass. 

By using the Gaia photometric data, we applied the MiMO algorithm \citep{2022ApJ...930...44L} to fit the cluster's parameters. This algorithm constructs a mixture model of field stars and cluster members (both singles and binaries) to match the observed stellar number density distribution on the CMD. Only stars brighter than $G=16$ mag were included in the fitting, because stars fainter than this magnitude may exhibit significant $G_{\rm BP}$ or $G_{\rm RP}$ deviations from their true values \citep{2021A&A...649A...3R}, and the model may also have considerable deviations in the low-mass range \citep{2019A&A...632A.105C}. In practice, we assumed that the distance modulus of the Pleiades is fixed at $DM_{\rm c}=5.65$ mag (corresponding to the average modified parallax of kinematic members: $\varpi_{\rm c} =7.41$ mas), the metallicity is fixed at [Fe/H]=0.1~\citep{2022A&A...668A...4F}, and the binary mass ratio distribution follows a power law with exponent $\gamma_q=0$. Based on these assumptions, the best-fit values for the cluster age and extinction are $\log(Age/{\rm yr})=8.026$ and $A_V=0.135$ mag from the MiMO. Then, we downloaded the model magnitudes with these fixed or fitted parameters from the website of the PARSEC, and defined it as the fiducial photometric model for the Pleiades. It offers 476 sets of model magnitudes as functions of stellar mass from 0.09 to 5.2 $\mathcal{M}_{\odot}$. Among these, 112 sets in the range of 0.09 to 4.9 $\mathcal{M}_{\odot}$ were used in multiband fitting for stellar mass and mass ratio of cluster members (see Table~\ref{tab:ridgeline} and Section~\ref{BI}).

It should be noted that the choice of theoretical model and the accuracy of cluster parameters will not have a substantial impact on the subsequent construction of the empirical model. This is because the fiducial model obtained from the fitting will ultimately be adjusted to the real observational data of the cluster, whereas minor differences in models and parameters could be eliminated during the adjustment.

The fiducial model is represented by solid black lines in Figure \ref{fig:data_cmd}. It reveals that neither the optical nor the NIR model magnitudes can perfectly match the data. Particularly for stars with masses below 0.5$\mathcal{M}_{\odot}$ ($G \gtrsim 16$ mag), the optical model photometry is significantly redder than the observation, leading to a large amount of low-mass binaries may be misclassified as single stars. Small discrepancies exist throughout the rest of the higher mass range. They will also influence the determination of mass ratios to some extent. This mismatch highlights the necessity of correcting the discrepancies between observational data and the fiducial model, which will be discussed in the following section.

\section{Multiband Fitting Methodology and Results for the Pleiades Members} \label{sec:method}

Taking advantage of the large fraction of single stars in the cluster that perfectly conform to the main sequence of a SSP, we can quantify the discrepancy between the model and the observational data in terms of the stellar mass-magnitude relation. This allows us to correct the theoretical model and derive an empirical model for the cluster (Section~\ref{sec:correction}). Subsequently, based on this empirical model, we utilize the multiband fitting within a Bayesian framework to derive the posterior probability density distribution (PDF) of mass and mass ratio for single stars, or binaries (Section~\ref{sec:fit}).

\subsection{Correction of Photometric Model and Construction of Empirical Isochrone}\label{sec:correction}

\subsubsection{Correction Strategy}\label{sec:single fit}

In studies of photometric binaries in clusters, empirical models can be employed as a more practical substitute for theoretical ones. For instance, in works based on CMD, a typical procedure is to hold the magnitude constant and adjust the model in the colors, to match the observed main sequence of single stars, i.e. using a $m-\Delta C$ relation to calibrate the model isochrone, where $\Delta C$ is the difference between the observed and model colors \citep{2019A&A...622A.110F,2012A&A...540A..16M,2020ApJ...901...49L}.

\begin{figure*}
	\includegraphics[width=\textwidth]{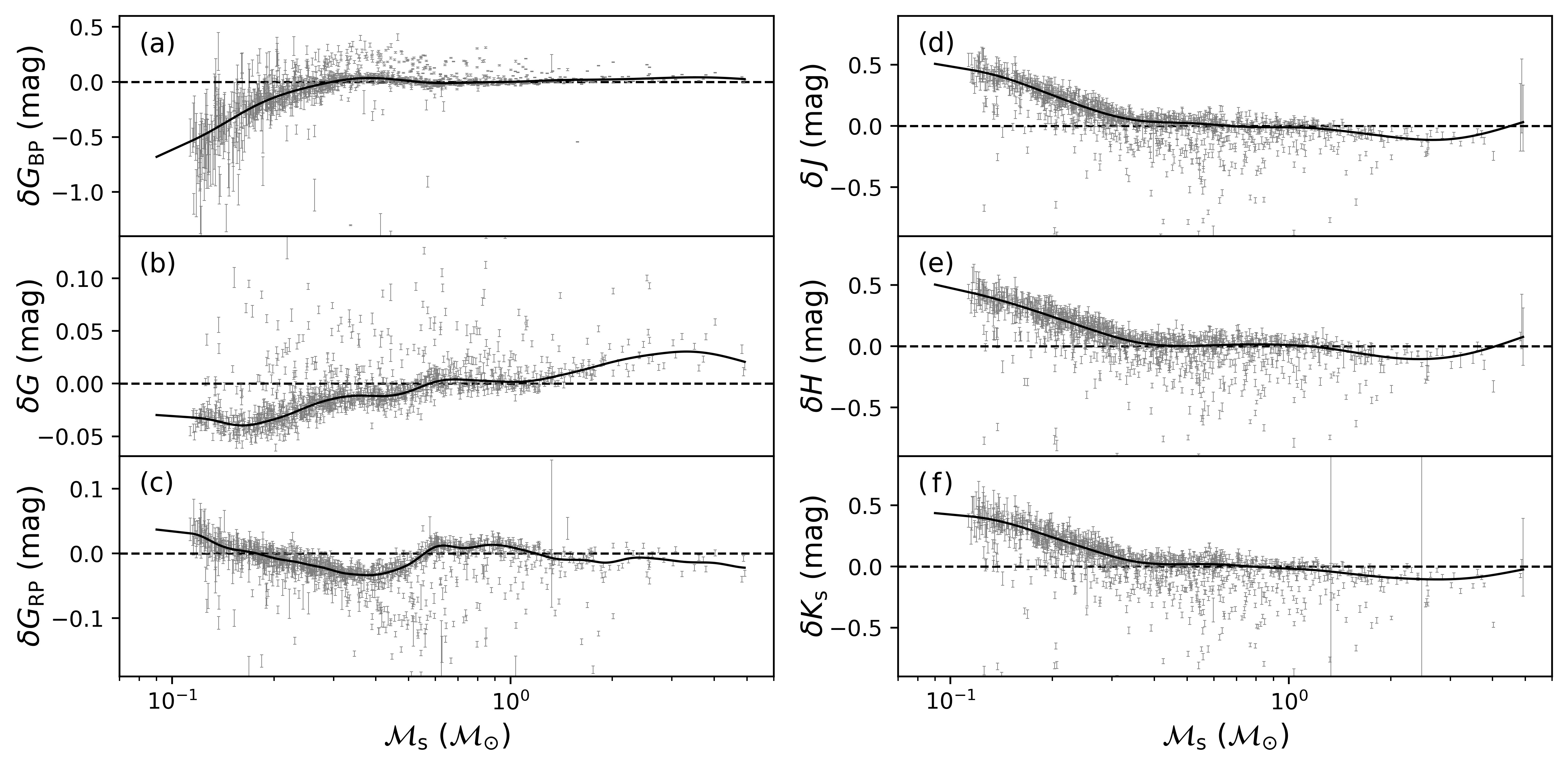}
    \caption{$\delta m - \mathcal{M}_{\rm s}$ planes and ridgelines of six photometric bands for the Pleiades. $\mathcal{M}_{\rm s}$ represents the mass derived from the single-star Multiband fitting. $\delta m$ represents the magnitude difference between the observational data and the best-fit model in each band. Each error bar corresponds to an individual star, with the length indicating the observational error. The solid black line shows the ridgeline that represents the $\mathcal{M}_s - \Delta m$ relation. The dashed black line marks the position where the magnitude difference equals zero.}
    \label{fig:ridgeline}
\end{figure*}

\begin{table*}
    \caption{The Fiducial Photometric Model and Its Correction $ \Delta m (\mathcal{M}_{\rm s})$ of the Gaia and 2MASS Bands for the Pleiades.}
    \label{tab:ridgeline}
    \vspace{-10pt}
    \begin{center}
    \setlength{\tabcolsep}{7.35pt}
    \begin{tabular}{ccccccccccccc}
\hline
\hline
$\mathcal{M}_{\rm{s}}$ & $ G_{\rm{BP}}$ & $ G$ & $G_{\rm{RP}}$ & $J$ & $H$ & $K_{\rm{s}}$ & $\Delta G_{\rm{BP}}$ & $\Delta G$ & $\Delta G_{\rm{RP}}$ & $\Delta J$ & $\Delta H$ & $\Delta K_{\rm{s}}$ \\
$(\mathcal{M}_{\odot})$ & (mag) & (mag) & (mag) & (mag) & (mag) & (mag) & (mag) & (mag) & (mag) & (mag) & (mag) & (mag) \\
(1) & (2) & (3) & (4) & (5)  & (6)  & (7) & (8) & (9) &(10) & (11) & (12) & (13) \\
\hline
0.0900 & 16.667 & 13.883 & 12.400 & 9.635 & 9.043 & 8.719 & -0.680 & -0.030 & 0.037 & 0.508 & 0.504 & 0.436 \\
0.0924 & 16.620 & 13.835 & 12.352 & 9.587 & 8.995 & 8.671 & -0.664 & -0.030 & 0.036 & 0.503 & 0.497 & 0.433 \\
0.1000 & 16.469 & 13.682 & 12.199 & 9.434 & 8.842 & 8.519 & -0.616 & -0.031 & 0.034 & 0.487 & 0.476 & 0.424 \\
0.1080 & 16.328 & 13.539 & 12.056 & 9.292 & 8.698 & 8.376 & -0.569 & -0.031 & 0.032 & 0.471 & 0.455 & 0.416 \\
0.1196 & 15.990 & 13.280 & 11.811 & 9.106 & 8.512 & 8.192 & -0.505 & -0.033 & 0.028 & 0.450 & 0.427 & 0.403 \\
0.1200 & 15.973 & 13.269 & 11.801 & 9.100 & 8.507 & 8.187 & -0.503 & -0.033 & 0.028 & 0.449 & 0.426 & 0.403 \\
0.1366 & 15.369 & 12.871 & 11.441 & 8.895 & 8.300 & 7.987 & -0.409 & -0.036 & 0.014 & 0.413 & 0.385 & 0.376 \\
0.1400 & 15.258 & 12.795 & 11.372 & 8.854 & 8.259 & 7.947 & -0.390 & -0.037 & 0.012 & 0.404 & 0.377 & 0.369 \\
0.1593 & 14.756 & 12.445 & 11.052 & 8.654 & 8.057 & 7.752 & -0.289 & -0.040 & 0.005 & 0.355 & 0.331 & 0.328 \\
0.1600 & 14.738 & 12.433 & 11.040 & 8.647 & 8.050 & 7.745 & -0.286 & -0.040 & 0.004 & 0.353 & 0.329 & 0.326 \\
\hline
\end{tabular}

    \end{center}
    \tablecomments{Columns (1)-(7) are the initial stellar mass ($\mathcal{M}_s$) and its corresponding magnitudes of six photometric bands from the PARSEC1.2s with [Fe/H]=0.1, $\log(Age)/yr=8.026$ and $A_{\rm V}=0.135$ mag.   Columns (8)-(13) show the corresponding correction values for each band. The empirical model magnitudes are expressed as $M_{\rm{emp}}(\mathcal{M}_{\rm s})= M_{\rm{model}} (\mathcal{M}_{\rm s}) + \Delta m(\mathcal{M}_{\rm s}) $.   The complete version of this table is available in a machine-readable format.}
\end{table*}

In the present work, we propose an optimized correction method aiming to directly establish relationships between the mass of single stars ($\mathcal{M}_{\rm s}$) and magnitude corrections  ($\Delta m$) in various bands, rather than $m-\Delta C$ relations. Subsequently, we adjust the theoretical model to an empirical model for stellar magnitudes, namely, a set of $M_{\rm emp}(\mathcal{M}_{\rm s})$ functions of corresponding bands. The steps involved are as follows.

We first assume that all cluster members are single stars, and used the fiducial model for a multiband fitting of the stellar mass for each member. For the $i$th target, we employed the standard $\chi^2$ fit to produce its best-fit mass ($\mathcal{M}_{\rm s}$) and corresponding model magnitudes in each band. Subsequently, we calculated the discrepancy between the best-fit model and the observed magnitudes, defined as $\delta m(i)=m_{\rm obs}(i)-M_{\rm model}(i) -DM_c$. Then, $\delta m$  of all stars with their stellar masses can be obtained as plotted in panels of Figure~\ref{fig:ridgeline} for each band.

Ideally, if both the observational and model magnitudes were exactly correct, and each cluster member were a single star, then all $\delta m$ values would equal 0 mag. In the case of the theoretical model not aligning perfectly with the observational data, single stars will appear around a curve that deviates from the horizontal line on the $\delta m -\mathcal{M}_{\rm s}$ plane but remain highly concentrated. High mass-ratio binaries in the cluster and some contaminated field stars, which do not fit well with the single-star model, will distribute in a more scattered region.

Finally, we employed a new robust regression method based on the standard Gaussian Process (GP) and Iterative Trimming (ITGP) algorithm \citep{2021A&C....3600483L} to determine the ridgeline on the $\delta m - \mathcal{M}_{\rm s}$ plane,  where the main-sequence single stars should converge. ITGP is specially designed to rule out outliers through an iterative process, and has been successfully applied to model the main sequence on CMDs of open clusters \citep{2020ApJ...901...49L}. The ridgelines, shown as solid lines in Figure~\ref{fig:ridgeline} represent the $\Delta m(\mathcal{M}_{\rm s})$ functions for each band. They were extrapolated to the lower mass limit of the PARSEC model (0.09$\mathcal{M}_{\odot}$) to obtain complete ones.

\subsubsection{Empirical Model of the Pleiades' Photometry}\label{sec:relation}

As shown in Figure~\ref{fig:ridgeline}, for stars with $0.6\mathcal{M}_{\odot} < \mathcal{M}_{\rm s} < 1\mathcal{M}_{\odot}$, the magnitude corrections $\Delta m (\mathcal{M}_{\rm s})$  are relatively small across all bands. While larger corrections are observed in both lower and higher mass ranges. For example, in the low-mass range, the maximum corrections are approximately 0.03 mag for $G$ and $G_{\rm{RP}}$, 0.07 mag for $G_{\rm BP}$ and up to 0.5 mag for, $J, H$ and $K_{\rm{s}}$. 

Table~\ref{tab:ridgeline} presents the original PARSEC model magnitudes alongside their corresponding correction values, for stellar masses ranging from 0.09 to 4.9$M_{\odot}$. The empirical model is then derived as $M_{\rm{emp}}(\mathcal{M}_{\rm s})= M_{\rm{model}} (\mathcal{M}_{\rm s}) + \Delta m(\mathcal{M}_{\rm s})$ for each band. It is worth noting that this calibration is made-to-measure for the Pleiades cluster with its fundamental parameters. That means the magnitude corrections not only depend on stellar mass but also exhibit slight variations with the adopted metallicity, extinction, and age of the cluster. Therefore, when applying these correction values to other clusters with different cluster parameters, minor deviations should be anticipated.

\subsection{Multiband Fitting for Stellar Mass and Mass Ratio of Cluster Members}\label{sec:fit}

We employed the empirical model constructed from Gaia and 2MASS magnitudes to perform multiband fitting on members of the Pleiades. This enabled us to determine their stellar masses, including those of single stars and the primary and secondary components of binaries. Our fitting procedure, conducted within a Bayesian inference framework, introduced an optimized form of the model magnitudes of an individual object in the cluster. The posterior PDF of fitting parameters was then derived for each member, allowing us to calculate their best estimates and uncertainties. Additionally, we computed the probability of each member being a binary, denoted as $P_{\rm b}$.

\subsubsection{Model Magnitudes of Cluster Members and the Lower Mass Limitation}\label{sec:binary fit}

Let us assume that each cluster member, which is an unresolved object, possesses a companion, and define a photometric binary to be an object with detectable luminosity contribution from its secondary star. In this manner, we can adopt a unified stellar magnitude expression of the object. For any given band, the absolute magnitude in the empirical model can be expressed as follows:
\begin{equation}
    M'_{\rm emp} = 
    \begin{cases}
        \mathord{-}2.5 \log[10^{\mathord{-}0.4 M_{\rm emp}(\mathcal{M}_1)} + 10^{\mathord{-}0.4 M_{\rm emp}(q\mathcal{M}_1)}]\\\hspace{1.8cm} &  \hspace{-3.5cm} q\mathcal{M}_1 \geq \mathcal{M}_{\rm lim}, \\
        M_{\rm emp}(\mathcal{M}_1), &  
        \hspace{-3.5cm} q\mathcal{M}_1 <  \mathcal{M}_{\rm lim},
    \end{cases}
    \label{eq:mb}
\end{equation}
where $M_{\rm emp}(\mathcal{M}_1)$ and $M_{\rm emp}(q\mathcal{M}_1)$ represent the absolute magnitudes of the primary and secondary stars from the empirical model, respectively. Then the modeled apparent magnitude of this object is 
\begin{equation}
    m_{\rm emp} = M'_{\rm emp} +DM 
    \label{eq:mb_emp}
\end{equation}
with $DM$ to be its distance modulus.

$\mathcal{M}_{\rm lim}$ denotes the lowest stellar mass in the model. Our empirical model is modified from the PARSEC 1.2s model, which has a lower mass limit of 0.09$\mathcal{M}_{\odot}$, slightly higher than the well-known stellar mass limit of $\sim 0.075 \mathcal{M}_{\odot}$. Equation~(\ref{eq:mb}) implies an assumption that companion stars with masses below 0.09$\mathcal{M}_{\odot}$ are too faint to contribute a detectable luminosity increment. So, they are considered, or defined as single stars in this work. 

\begin{figure}
    \centering
	\includegraphics[width=0.94\columnwidth]{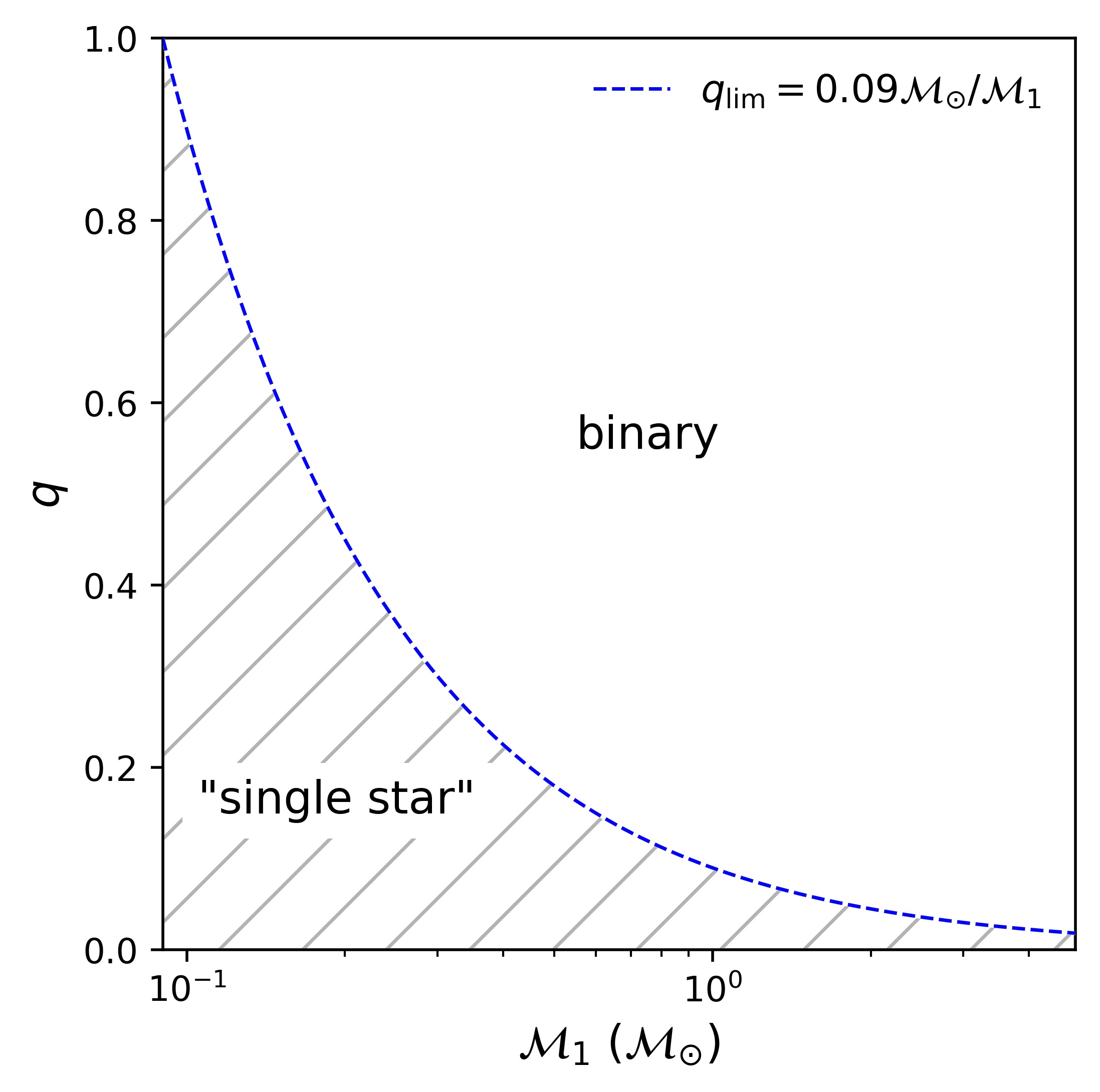}
    \caption{Lower limit of the detectable mass ratio ($q_{\rm lim}$) varying with primary mass. The blue line denotes the threshold where the secondary mass is 0.09$\mathcal{M}_{\odot}$. Stars with mass ratios above this line are binary stars, and those below are considered ``single stars''.}
    \label{fig:method}
\end{figure}

The secondary mass threshold results in a lower limit on detectable binary mass ratios, which is a function of the primary mass, $q_{\rm{lim}}(\mathcal{M}_1)=\mathcal{M}_{\rm lim}/\mathcal{M}_1$, as shown by the blue line in Figure~\ref{fig:method}. This line separates the $q-\mathcal{M}_1$ plane into two distinct regions. Stars above this line are classified as binaries, whereas those below are considered as single stars, regardless of their $q$ values.

\subsubsection{Bayesian Inference}\label{BI}

For each cluster member, its age, metallicity, and extinction are fixed at the cluster's values, and the distance modulus is fixed at its own modified $DM$ (see Appendix~\ref{apen:BayesPlx} and Table~\ref{tab:member} for details). Then we have two parameters, the primary mass $\mathcal{M}_1$ and the mass ratio $q$ need to be fit.

In the multiband fitting, a formal $\chi^2$ is written as: 
\begin{equation}
 \chi^2 = \sum_j \chi_j^2= \sum_j \frac{(m_{{\rm obs},j}-m_{{\rm emp},j})^2}{err_j^2} , 
    \label{eq:chisq}
\end{equation}
where $j$ denotes six observational bands, including $G_{\rm{BP}}$, $G$, $G_{\rm{RP}}$, $J$, $H$, and $K_{\rm{s}}$, $m_{{\rm obs},j}$ represents observed magnitude in each band, $err_j$ indicates the observational error involving error floors of 0.01mag for the Gaia data, and $m_{{\rm emp},j}$ refers to the empirical model magnitude expressed by Equations~(\ref{eq:mb}) and (\ref{eq:mb_emp}).

In the Bayesian Framework, we often transfer the $\chi^2$ to the likelihood function, which is a probability for obtaining the observational magnitudes at given $\mathcal{M}_1$ and $q$ and is defined as:
\begin{equation}
    \mathcal{L}=\exp(-\chi^2/2)
    \label{eq:likelihood}
\end{equation}
It is certain that the maximum likelihood is corresponding to the least $\chi^2$  point. Besides, the advantage of using the Bayesian Inference is that it not only focuses on the best-fit points, but also produces an entire posterior PDF in the fitting-parameter space.

For priors of fitting parameters, we set $\mathcal{M}_1$ follows a log-uniform distribution ranging from 0.09 to $4.9\mathcal{M}_{\odot}$, and $q$ follows a uniform distribution between 0 and 1. We employed the Nautilus \citep{nautilus} package, which uses an importance-sampling technique, to generate the posterior PDF for each star in the parameter space. During the sampling of $\mathcal{M}_1$ and $q$, the value of $M_{\rm emp}$ is linearly interpolated between model values listed in Table~\ref{tab:ridgeline}.

\begin{table*}
    \caption{Results of Multiband Fitting for Stars with $P_{\rm{k}}\geq0.5$}
    \vspace{-10pt}
    \label{tab:result}
    \begin{center}
    \setlength{\tabcolsep}{8.3pt}
    \begin{tabular}{ccccccccccc}
\hline
\hline
 SourceId&  $\mathcal{M}_{\rm{1}}$  & $q$ &  $\mathcal{M}_{\rm{1,16}}$ &  $\mathcal{M}_{\rm{1,50}}$ &  $\mathcal{M}_{\rm{1,84}}$ & $q_{16}$ & $q_{50}$ & $q_{84}$ &   $P_{\rm{b}}$ &  flag of exclusion\\
 &  ($\mathcal{M}_{\odot}$) &  & ($\mathcal{M}_{\odot}$) &  ($\mathcal{M}_{\odot}$)  & ($\mathcal{M}_{\odot}$) &&&&&\\
(1) & (2) & (3) & (4) & (5) & (6) & (7) & (8) & (9) & (10) & (11)\\
\hline
68051390279853824 & 0.656 & 0.11 & 0.655 & 0.656 & 0.657 & 0.02 & 0.07 & 0.12 & 0.000 & ... \\
70461554129560448 & 0.662 & 0.14 & 0.661 & 0.662 & 0.663 & 0.04 & 0.13 & 0.18 & 0.495 & ... \\
71377000637253504 & 0.659 & 0.38 & 0.658 & 0.659 & 0.661 & 0.35 & 0.38 & 0.41 & 1.000 & ... \\
68587161680620288 & 0.571 & 1.00 & 0.574 & 0.579 & 0.586 & 0.95 & 0.97 & 0.99 & 1.000 & ... \\
64401389632616960 & ... & ...  & ... & ... & ... & ... & ... & ... & ... & $K_{\rm{s}}$ \\
51761575759616000 & ... & ...  & ... & ... & ... & ... & ... & ... & ... & $G_{\rm{BP}}$ \\
66167411464494848 & ... & ...  & ... & ... & ... & ... & ... & ... & ... & close neighbor \\
\hline
\end{tabular}

    \end{center}
    \tablecomments{The SourceId is the identifier provided by Gaia DR3. Columns (2) and (3) list the best-fit values of  $\mathcal{M}_1$ and $q$. Columns (4)-(9) provide the [16$\%$, 50$\%$, 84$\%$] quantiles of these two parameters. The probability of a star being a binary ($P_{\rm{b}}$) is derived from the probability density function (PDF). The ``flag of exclusion'' indicates the criteria used for excluding field stars and non-main-sequence members. (The complete version of this table is available in a machine-readable format.)
}
\end{table*}

With the posterior PDF derived from the sampling process, we can further quantify the probability of each star being a binary system. Specifically, using the posterior PDF of the $i$th object in the $q-\mathcal{M}_1$ plane, denoted as $p_i (\mathcal{M}_1,q)$, we can determine the binary probability of this target, $P_{\rm b}(i)$, as the proportion of the PDF that lies above $q_{\rm{lim}}$ (illustrated by the blue line in Figure~\ref{fig:method}):
\begin{equation}
    P_{\rm{b}}(i)= \frac{ \displaystyle\int_{\mathcal{M}_1}\int_{q>q_{\rm lim}} p_i(\mathcal{M}_1,q) d\mathcal{M}_1 dq}  { \displaystyle\int_{\mathcal{M}_1} \int_{q} p_i(\mathcal{M}_1,q) d\mathcal{M}_1 dq}.
    \label{eq:pb}
\end{equation}

\subsubsection{Fitting Result of the Pleiades}\label{sec:result}

After removing 153 stars with neighbors in 5 arcsec, we performed multiband fitting on the remaining 1237 kinematic members. The typical values of fitted parameters can be characterized by the best-fit values, where the maximum posterior probability is located, or the median value of each parameter from its corresponding marginal PDF. The uncertainties can be estimated as half of the [16$\%$, 84$\%$] interval of the marginal PDFs.  The fitting results show high precision, especially for $\mathcal{M}_1$. For stars with $P_{\rm b}>0.5$, most of $\mathcal{M}_1$ uncertainties are smaller than 0.01$\mathcal{M}_{\odot}$. The precision of $\mathcal{M}_1$ may be even higher for single stars. This is mainly due to the high precision of photometric observations. For instance, the value of $d\mathcal{M}/d G$ is only about 0.15$\mathcal{M}_{\odot}$ mag$^{-1}$ for a star of one solar mass, while the typical error of $G$ is 0.01mag. So the precision of $\mathcal{M}_1$ will be better than 0.01$\mathcal{M}_{\odot}$ for single stars. The joint usage of six magnitudes will further improve the precision. Uncertainties of $q$ for most stars with large $P_{\rm b}$ are smaller than 0.1.

These fitting results, together with the binary probabilities are listed in Table~\ref{tab:result}. For total 1237 fitted objects, 83 were further excluded due to their poor fits, characterized by $\chi_j^2 > 25$ in any band, indicating significant discrepancies between the empirical model and observation. Detailed exclusion reasons are also provided in Table~\ref{tab:result}, column ``flag of exclusion''. These excluded stars may be MS+WD binaries or multiple-star systems in the cluster, or possibly some contaminated background stars. In brief, the multiband fitting can be regarded as an independent approach to diagnose the main-sequence cluster members, which can rule out those stars that do not satisfy the criteria for a single MS star or a MS+MS binary within the Pleiades. Then the remaining sample of 1154 main-sequence members will be analyzed in detail for the binaries. Notably, this is not a complete sample of multiple systems in the Pleiades, since it only contains the MS+MS binaries, and does not include stars with only 2-parameter solutions, which account for just 1$\%$ of the total sample.

\begin{figure*}
	\includegraphics[width=\textwidth]{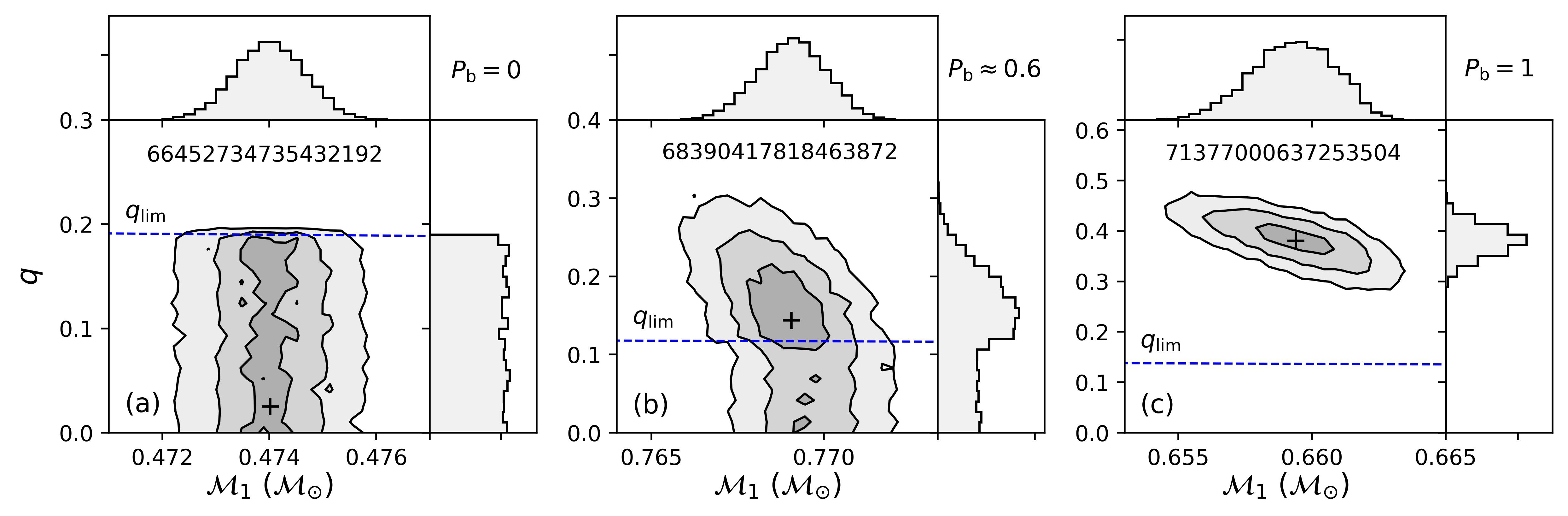}
    \centering
    \caption{Probability density functions (PDFs) of three members. The contour plots in panels (a), (b), (c) illustrate the PDFs of stars with different binary probabilities: $P_{\rm{b}}=0,0.6,1$, respectively. The SourceIds of Gaia DR3 are marked in each panel. Each contour corresponds to the $1\sigma, 2\sigma, 3\sigma$ (39.3$\%$, 86.5$\%$ and 98.9$\%$) confidence levels. The blue lines denote the lower limit of detectable mass ratios.}
    \label{fig:example}
\end{figure*}

Figure~\ref{fig:example} presents posterior PDFs for three representative objects: a single star (panel (a)), a binary (panel (c)), and an intermediate case (panel (b)). The plus symbols indicate the best-fit points. In the single star part (below the $q_{\rm lim}$ in panels (a) and (b)), the PDF is approximately parallel to the $q$ coordinate. It indicates that there is no constraint on $q$, which is a natural consequence of the form of Equation~(\ref{eq:mb}). One should also notice that the best-fit value of $q$ could be randomly assigned for a single star, since we really do not know the mass of the potentially existing ``dark'' component. In the binary parts of the PDFs (above the $q_{\rm lim}$ in panels (b) and (c)), a significant degeneracy can be found between $q$ and $\mathcal{M}_1$, indicating that during the fitting process, the reduction in the primary mass needs to be compensated by a larger secondary star. Panel (b) is a good example to show the advantage of the Bayesian inference with the posterior PDF. If we only consider the best fit, this object would be classified as a binary. But in fact, it has approximately 40 percent to be a single star, which is revealed by its PDF.

\begin{figure}
    \centering
	\includegraphics[width=0.94\columnwidth]{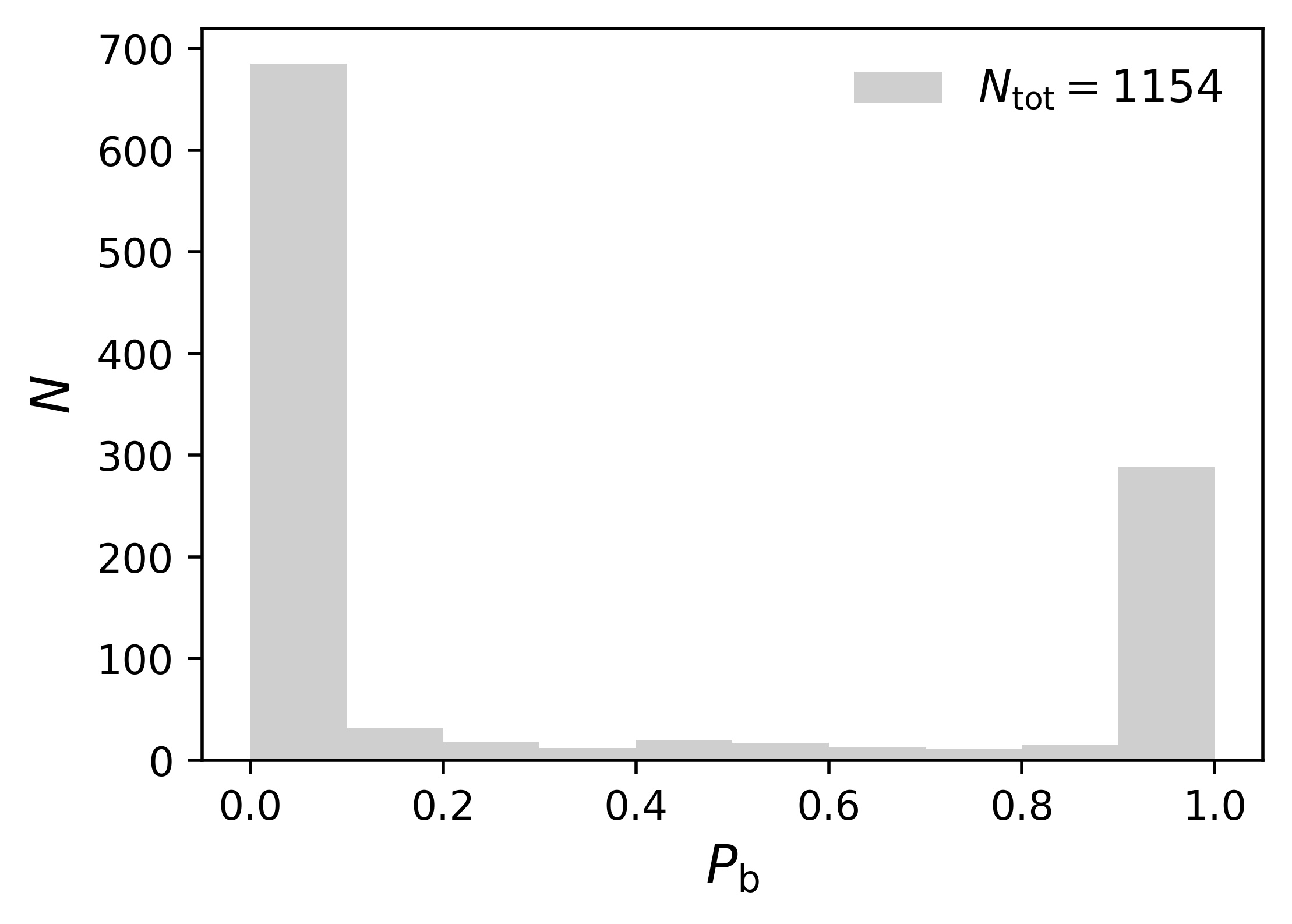}
    \caption{Distribution of binary probability ($P_{\rm b}$) of main-sequence members.}
    \label{fig:pb_1}
\end{figure}

Figure~\ref{fig:pb_1} shows the distribution of $P_{\rm b}$ for the final main-sequence sample, indicating that most members are concentrated in the regions of $P_{\rm{b}} < 0.1$ or $P_{\rm{b}} > 0.9$. The effectiveness index$^1$ of $P_{\rm b}$ is $E=0.88$, representing an excellent separation. This result demonstrates that combining Gaia and 2MASS photometry can effectively distinguish binaries from single stars in the Pleiades.

\begin{figure*}
	\includegraphics[width=\textwidth]{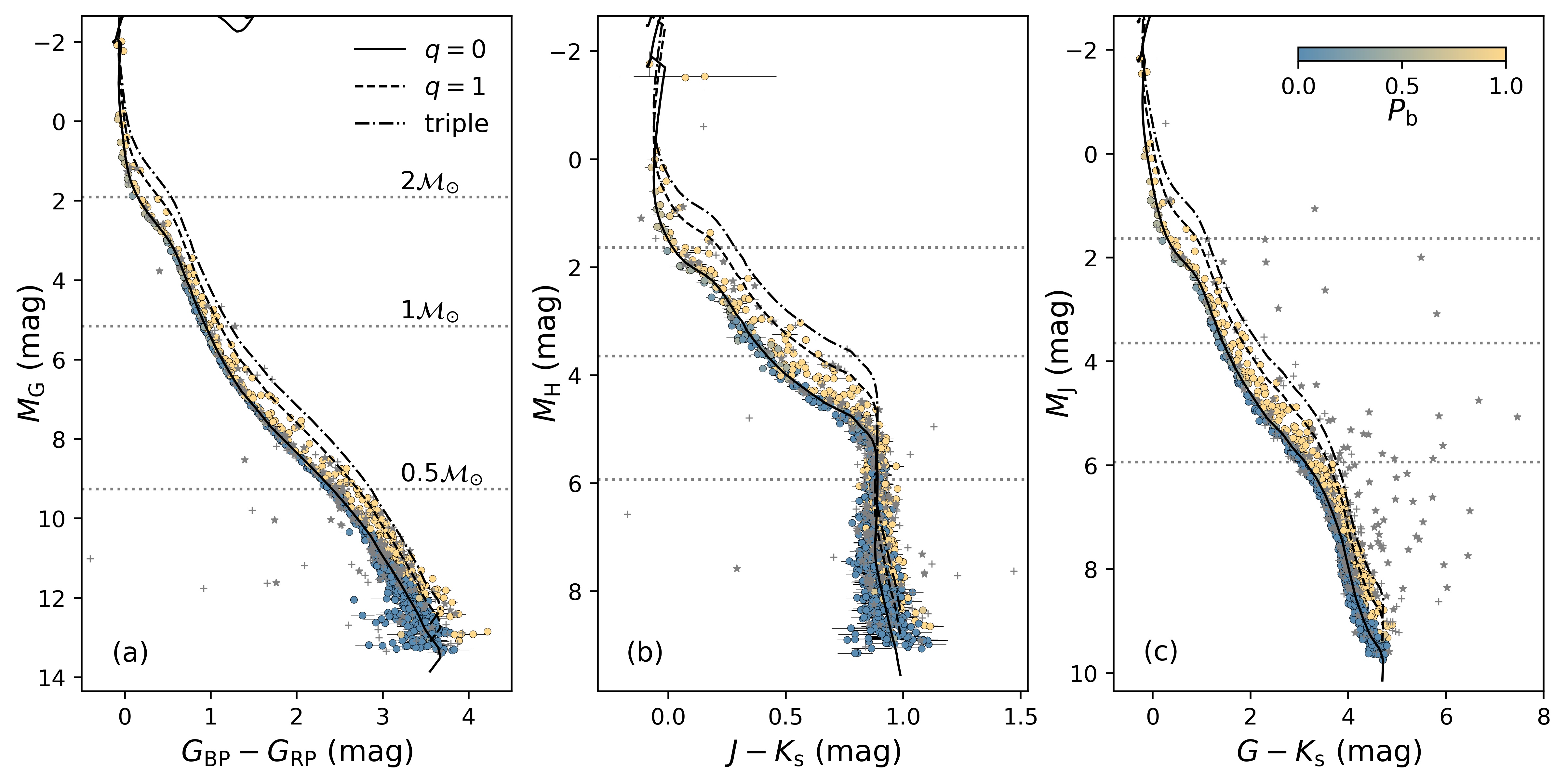}
    \caption{Distributions of binary probability ($P_{\rm b}$) for Pleiades kinematic members on CMDs of Gaia, 2MASS, and the combined data. The star symbols mark the stars with neighbors in 5 arcsec. The gray crosses show the stars excluded after multiband fitting. The colored points represent the final 1154 members, with color indicating $P_{\rm b}$. All photometric magnitudes have been adjusted to absolute magnitudes using the $DM$ in Table~\ref{tab:data}. The dotted lines mark $\mathcal{M}_{\rm s}=0.5, 1$ and $2\mathcal{M}_{\odot}$ from bottom to top. The black solid and dashed lines represent the empirical isochrones for $q=0$ and 1, respectively. The dot-dash lines represent isochrons for triple stars with equal masses.}
    \label{fig:pb}
\end{figure*}

Figure~\ref{fig:pb} shows the distribution of the final 1154 cluster members on the CMD, with their $P_{\rm{b}}$ values color-coded. Most stars exhibit $P_{\rm{b}}$ close to either 0 or 1, with only a few sources showing intermediate values. However, a lot of intermediate $P_{\rm{b}}$ values are found for massive stars, which implies that, by using current photometric data, it is still difficult to definitively determine whether or not they have a low-mass companion.

\begin{figure}
    \centering
	\includegraphics[width=0.94\columnwidth]{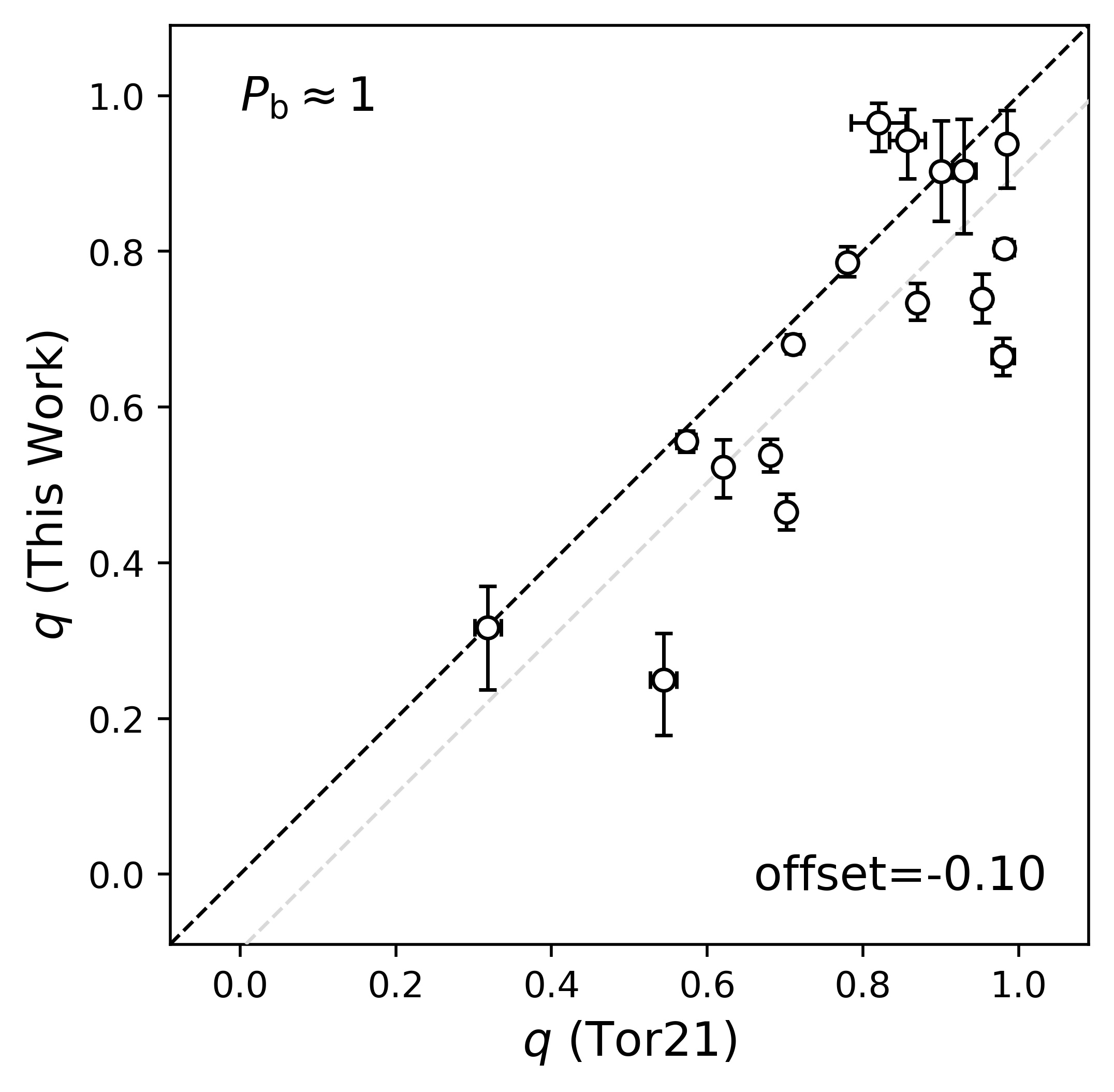}
    \caption{Comparison of mass ratio from this work with double-line binaries \citep{2021ApJ...921..117T}. All these stars have $P_{\rm b}\approx1$. The black line indicates y=x, and the gray line shows an offset of -0.10.}
    \label{fig:q_compare}
\end{figure}

We identified 17 common objects in our final sample and the dynamical mass-ratio subsample of \cite{2021ApJ...921..117T}. These stars are confirmed as photometric binaries in our work with $P_{\rm{b}} \sim 1$. Figure~\ref{fig:q_compare} shows the comparison of their $q$ values, revealing a significant correlation between these two independent measurements. This validates the reliability of our method for binary detection and the accuracy of the mass-ratio measurements. The binary with the lowest $q$ value in our results shows the largest discrepancy in $q$ between two $q$ values, but this divergence falls within the estimated uncertainties. Moreover, the $q$ values obtained from our photometric method are systematically smaller than those acquired from double-line spectroscopy, with a mean offset of -0.10. It indicates a systematic discrepancy between the two techniques. Similar conditions have been reported in other studies \citep{2020AJ....159...11C,2021AJ....161..160T,2024ApJ...962...41C}.

Additionally, it is interesting to mention that, since we have not used the RUWE value to constrain our kinematic member, our photometric diagnoses of binary results can be used in reverse to characterize the RUWE. For 685 single stars ($P_{b}<0.1$), all but two of them have RUWE$<$1.4. While for 331 binaries ($P_{b}>0.9$), 91 of them having RUWE$>$1.4. That means, using RUWE value as a criterion may only pick up one-third of binaries.

\section{Binary Fraction and Mass Ratio Distribution}\label{sec:discussion}

In this section, we generate combined probability density functions (PDFs) for all sample stars using a ``stacking'' method (Section~\ref{sec:stack}). These PDFs are then employed to analyze the binary fraction (Section~\ref{sec:fb}) and its correlation with the primary mass (Section~\ref{sec:fb_m}). We also investigate the mass-ratio distribution (Section~\ref{sec:qd}) and its dependence on the primary mass (Section~\ref{sec:qd_m}).

\subsection{Combined PDF of Primary Mass and Mass Ratio}\label{sec:stack}
\begin{figure*}
    \centering
	\includegraphics[width=\textwidth]{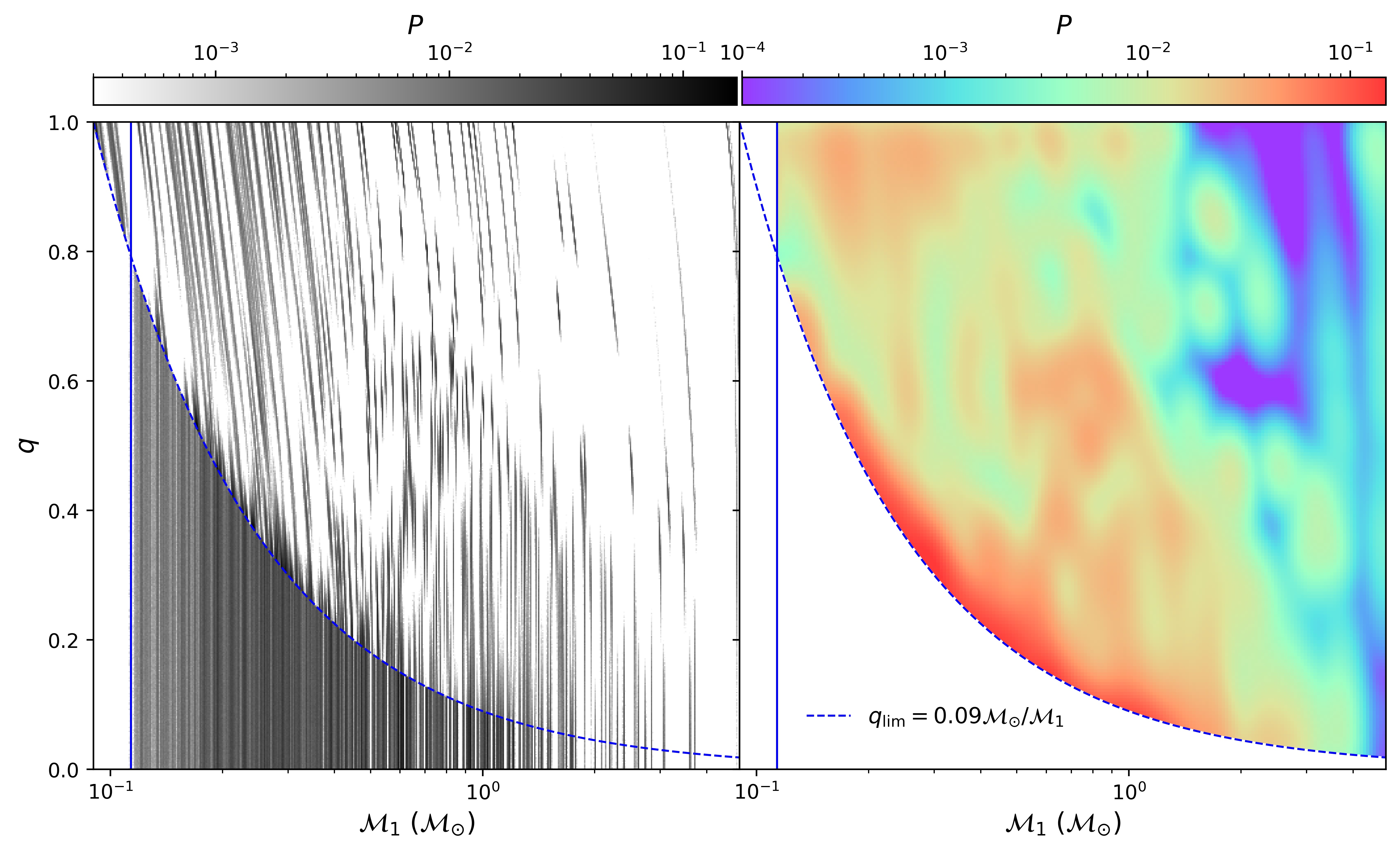}
    \caption{Combined probability density function (PDF) of primary mass and mass ratio for our sample. The left panel shows the direct result, with grayscale representing probability density. The dashed blue line indicates the position of the binary star with a secondary mass of 0.09 $\mathcal{M}_{\odot}$. Binaries are located above this line, and single stars are below it. The solid blue line marks $\mathcal{M}_1=0.11$. The stars to the left of this line are incomplete and are thus excluded from the statistical analysis. The right panel presents a smoothed version of the left panel, displaying only the region for binary analysis, with color indicating probability density.}
    \label{fig:result}
\end{figure*}

In general, the best-fit or median values of $\mathcal{M}_1$ and $q$ for each member star in Table~\ref{tab:result} can be used for the statistical analysis of binaries in the cluster. However, from the Bayesian perspective, this method lacks rigor and may be susceptible to increased statistical fluctuations, particularly when the number of a sample or sub-sample is limited. To address this issue, we propose a novel approach. As shown in the left panel of Figure~\ref{fig:result}, we stacked the normalized PDFs for all sample stars in the $q-\mathcal{M}_1$ plane to establish a combined PDF of the main-sequence stars in the cluster, denoted as $P(\mathcal{M}_1,q)$, where the capital $P$ represents probability density for the whole sample, distinguishing it from that of individual stars, $p_i$. The $P(\mathcal{M}_1,q)$ satisfies the following equation:
\begin{equation}
\displaystyle\int_{\mathcal{M}_1} \int_q P(\mathcal{M}_1,q) d\mathcal{M}_1dq = N ,
    \label{eq:Ntot}
\end{equation}
\noindent where $N=1154$ represents the number of photometric members in our final sample.  The $P(\mathcal{M}_1,q)$ essentially acts as a complete ``mapping'' from the photometric data to the stellar mass distribution of the cluster, which encompasses all information about the single, primary, and secondary stars.

In our sample, the lower luminosity limit is $G=19$ mag, corresponding to a single star mass of $0.11 \mathcal{M}_{\odot}$. Consequently, the following binary analysis is limited to $\mathcal{M}_1>0.11\mathcal{M}_{\odot}$, which is to the right of the blue solid vertical line in Figure~\ref{fig:result}. The total number of stars in this region, derived from integrating the PDF, is $n=1145.2$. Note that this value is a fractional number from the integration rather than a direct counting. We use $n$ to distinguish it from the simple count of stars, $N$.

When we use $P(\mathcal{M}_1,q)$ to investigate the binary fraction and its relation with primary mass and mass ratio, it is flexible to calculate various statistical parameters by using integration instead of counting in all cases. For example, the total number of binaries in our sample with primary masses higher than 0.11$\mathcal{M}_{\odot}$ is
\begin{equation}
n_{\rm b}=\displaystyle\int_{\mathcal{M}_1>0.11\mathcal{M}_{\odot}} \int_{q\ge q_{\rm lim}} P(\mathcal{M}_1,q) d\mathcal{M}_1 dq. 
    \label{eq:Nb}
\end{equation}
\noindent Although $n_{\rm b}$ is the result of integration, it is still subject to the ``law of small numbers", so its uncertainty can be roughly estimated by $n_{\rm b}^{1/2}$ based on the Poisson fluctuation. Thus, the total binary fraction of our sample is $f_{\rm{b}}= n_{\rm b} /n$, with a corresponding uncertainty of $n_{\rm b}^{1/2} /n$. Similarly, for other subsamples, such as when restricting the range of $\mathcal{M}_1$ or $q$, we can calculate the corresponding number of binaries, binary fraction, and their associated errors. Consequently, it will be easy to obtain $n_{\rm b}$ or $f_{\rm b}$ as functions of $\mathcal{M}_1$ and $q$.

\subsection{Binary Fraction}\label{sec:fb}
\begin{table*}
	\caption{Comparison of Binary Fraction with Other Works. }
    \vspace{-10pt}
	\label{tab:fb_compare}
    \begin{center}
    \setlength{\tabcolsep}{12.5pt}
	\begin{tabular}{cccccc}
\hline
\hline
This Work & \multicolumn{5}{c}{Previously Published Works} \\
\cline{2-6}
 $f_{\rm{b}}$ & $f_{\rm{b}}$ &Mass range ($\mathcal{M}_{\odot}$)&$q_{\rm{min}}$& Data & Reference\\
\hline
0.53 ± 0.04 & 0.25 ± 0.03 & 0.5 - 4.9 & - & $RV$ &\cite{2021ApJ...921..117T}\\
\hline
0.12 ± 0.02 & 0.14 ± 0.02 & 0.4 - 3.6 & 0.6 & $G_{\rm{BP}},G,G_{\rm{RP}}$ & \cite{{2021AJ....162..264J}}\\
0.12 ± 0.02 & 0.059 ± 0.040 & 0.3 - 0.9 & 0.6 & $G_{\rm{BP}},G,G_{\rm{RP}}$ & \cite{{2023AA...672A..29C}}\\
0.11 ± 0.01 & 0.078 ± 0.009 & 0.180 - 2.295 & 0.6 & $G_{\rm{BP}},G,G_{\rm{RP}}$ & \cite{{2023AA...675A..89D}} \\
0.22 ± 0.04 & 0.20 ± 0.03 & 0.6 - 1.0 & 0.5 & $G_{\rm{BP}},G,G_{\rm{RP}}$ & \cite{{2020ApJ...903...93N}}\\
0.54 ± 0.04 & 0.41 ± 0.04 & 0.57 - 3.75 & - & $G_{\rm{BP}},G,G_{\rm{RP}}$ & \cite{{2020ApJ...903...93N}}\\
0.34 ± 0.02 & $>0.19$ & All~(?) & - & $G_{\rm{BP}},G,G_{\rm{RP}}$ & \cite{{2023MNRAS.525.2315A}}\\
\hline
0.50 ± 0.04 & 0.70 ± 0.14 & 0.5 - 1.8 & - & $G_{\rm{BP}},H,K_{\rm{s}},W_1,W_2$ & \cite{{2022AJ....163..113M}}\\
0.50 ± 0.04 & 0.73 ± 0.03 & 0.5 - 1.8 & - & $G_{\rm{BP}},H,K_{\rm{s}},W_1,W_2$ & \cite{{2023AJ....165...45M}}\\
\hline
\end{tabular}

    \end{center}
    \tablecomments{Row one compares our results with studies based on radial velocity variations using spectroscopic observations. Rows two through seven compare our findings with studies based on Gaia optical photometry data. Rows eight and nine compare our results with studies combining Gaia's optical data with 2MASS and WISE infrared photometry data on a pseudo-colors diagram. The (-) denotes no restriction on the mass ratio range; hence, we use the total binary fraction for comparison.}
\end{table*}

This work covers a primary mass range from $0.11\mathcal{M}_{\odot}$ to $4.9\mathcal{M}_{\odot}$ for Pleiades members, representing the broadest mass range analyzed for binaries in this cluster to date. By using Equation~(\ref{eq:Nb}), we determine the total binary fraction of our sample $f_{\rm{b}} = 0.34 \pm 0.02$.

\begin{figure}
    \centering	
    \includegraphics[width=0.94\columnwidth]{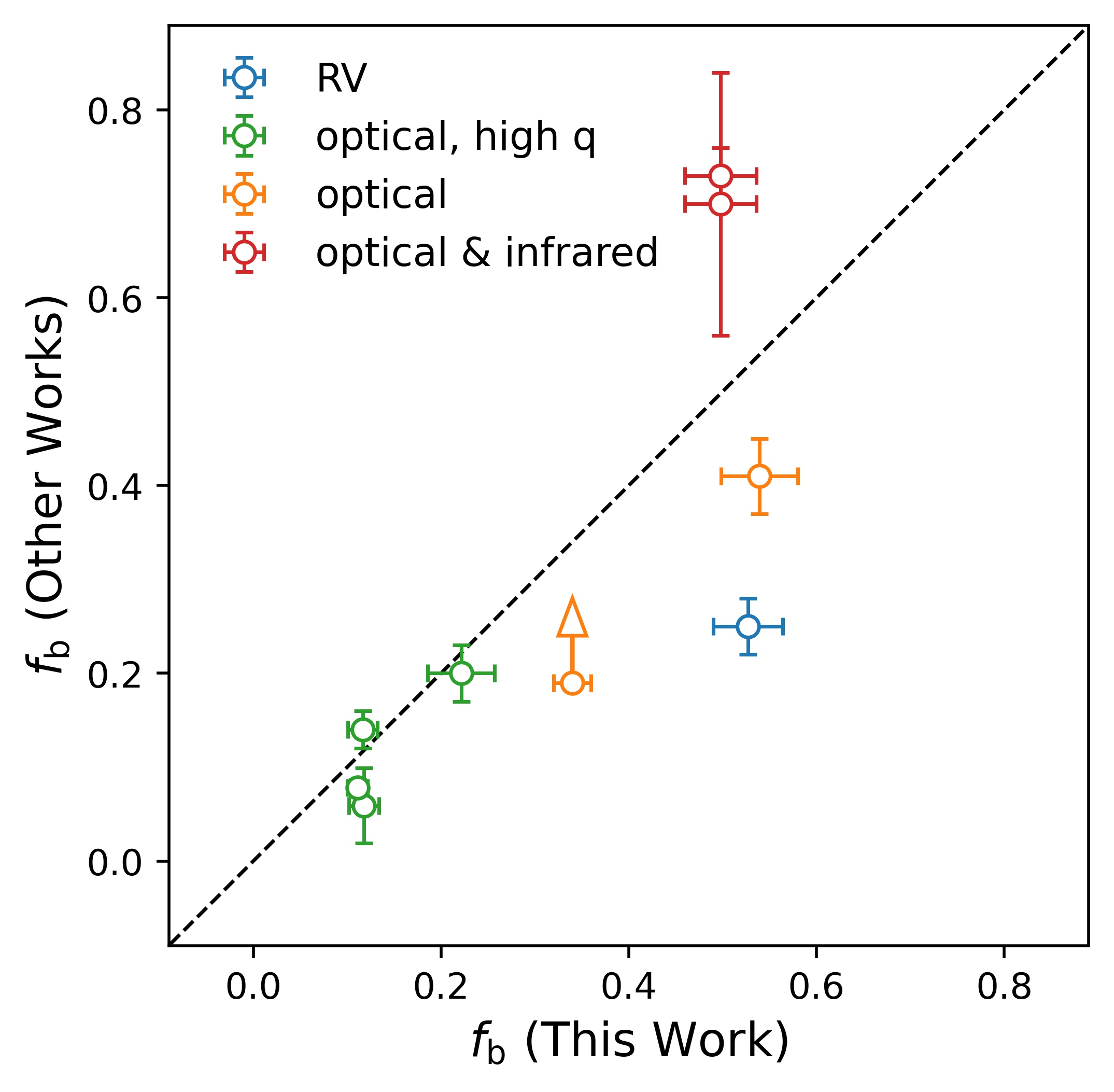}
    \caption{Comparison of binary fraction from this work with other works. The blue point indicates the comparison with a result using spectroscopic observations. Orange and green points represent comparisons with results using Gaia photometric data, with green points specifically showing high mass-ratio binary fraction. Red points show comparisons with results that combined Gaia, 2MASS, and WISE photometric data. The black line indicates y=x.}
    \label{fig:fb_compare}
\end{figure}

Numerous studies have investigated $f_{\rm b}$ of the Pleiades, with the range of mass and mass ratio varying depending on observations and methodologies. To compare with previous studies, we calculate $f_{\rm b}$ separately within the same mass and mass-ratio ranges of them. These results are detailed in Table~\ref{tab:fb_compare} and visually compared in Figure~\ref{fig:fb_compare}. Depending on the type of observational data used, these studies can be divided into three categories.

The first one is the spectroscopic binaries by using radial velocity variations \citep{2021ApJ...921..117T}. They found $f_{\rm b}=0.25 \pm 0.03 $ for periods shorter than 10000 days, in a $G$ magnitude range of $3.7-15$ mag, corresponding to a mass range of approximately $0.5-4.9\mathcal{M}_{\odot}$ for the Pleiades. This fraction is notably smaller than ours (see the blue point in Figure~\ref{fig:fb_compare}), which can be explained by the fact that the $\Delta RV$ method is less effective in detecting long-period, low mass-ratio binaries, or binaries with high orbital inclinations (face-on).
    
The second one is the photometric binaries by using optical data only. It includes several methods: separating regions of single and binary stars on CMDs with model lines \citep{2021AJ....162..264J,2023AA...672A..29C}, constructing a mixture model of single and binaries on CMDs \citep{2020ApJ...903...93N,2023AA...675A..89D}, or employing a kind of multiband fitting \citep{2023MNRAS.525.2315A}.  Our comparison reveals that the fractions of high mass-ratio binaries ($q > 0.6$ or 0.5) are consistent with ours (see green points in Figure~\ref{fig:fb_compare}). However, total binary fractions, including lower mass-ratio binaries, are significantly lower than ours (see orange points in Figure~\ref{fig:fb_compare}). This discrepancy supports our claim that using optical photometric data alone may miss a substantial number of low mass-ratio binaries, which will be more detectable in NIR bands.
    
At last, \cite{2022AJ....163..113M} and \cite{2023AJ....165...45M} combined the Gaia DR2, 2MASS, and WISE data and used the pseudo CMD of $(H-W_2)-W_1~vs~W_2-(G_{\rm{BP}}-K_{\rm{s}})$  to analyze binaries in the Pleiades. They divided the diagram into distinct mass ratio regions, and reported $f_{\rm b} =0.70\pm0.14$ and $0.73\pm0.03$, respectively, for the $\mathcal{M}_1$ range of $0.5-1.8\mathcal{M}_{\odot}$. These results are higher than ours, $f_{\rm b}=0.50\pm0.04$ (see red points in Figure~\ref{fig:fb_compare}). We propose that this discrepancy can be attributed to two factors: the inclusion of multiple-star systems in their sample and the contamination of single stars in the $q=0-0.2$ interval.

\begin{table*}
	\caption{Binary Fraction as a Function of Primary Mass for Different Mass Ratios in the Pleiades}
    \vspace{-10pt}
	\label{tab:fb_1}
    \begin{center}
    \setlength{\tabcolsep}{21.6pt}
	\begin{tabular}{cccccc}
\hline
\hline
$\mathcal{M}_1~(\mathcal{M}_{\odot})$ & $n$ & $f_{\rm{b}}$ & $f_{\rm{b}}^{0-0.3}$ & $f_{\rm{b}}^{0.3-0.8}$ & $f_{\rm{b}}^{0.8-1}$ \\
\hline
0.11 - 4.90 & 1145.2 &  0.34 ± 0.02 & 0.09 ± 0.01 & 0.17 ± 0.01 & 0.07 ± 0.01 \\
\hline
0.11 - 0.18 & 178.7 &  0.18 ± 0.03 & - & 0.08 ± 0.02 & 0.10 ± 0.02\\
0.18 - 0.30 & 345.7 &  0.23 ± 0.03 & - & 0.16 ± 0.02 & 0.07 ± 0.01\\
0.30 - 0.50 & 245.7 &  0.31 ± 0.04 & 0.10 ± 0.02 & 0.15 ± 0.02 & 0.06 ± 0.02\\
0.50 - 0.90 & 211.6 &  0.45 ± 0.05 & 0.15 ± 0.03 & 0.25 ± 0.03 & 0.05 ± 0.02\\
0.90 - 1.60 & 119.1 &  0.55 ± 0.07 & 0.24 ± 0.05 & 0.25 ± 0.05 & 0.06 ± 0.02\\
1.60 - 3.00 & 34.8 &  0.78 ± 0.15 & 0.51 ± 0.12 & 0.21 ± 0.08 & 0.06 ± 0.04\\
3.00 - 4.90 & 9.5 &  0.94 ± 0.31 & 0.32 ± 0.18 & 0.48 ± 0.23 & 0.13 ± 0.12\\
\hline
\end{tabular}

    \end{center}
    \tablecomments{$n$ denotes the number of stars within the specified mass range. $f_{\rm{b}}$ represents binary fraction for photometric detectable binaries with $q>q_{\rm{lim}}(\mathcal{M}_1)$. The (-) denotes a range where photometric binaries cannot be detected. }
\end{table*}

\begin{figure*}
    \centering
	\includegraphics[width=\textwidth]{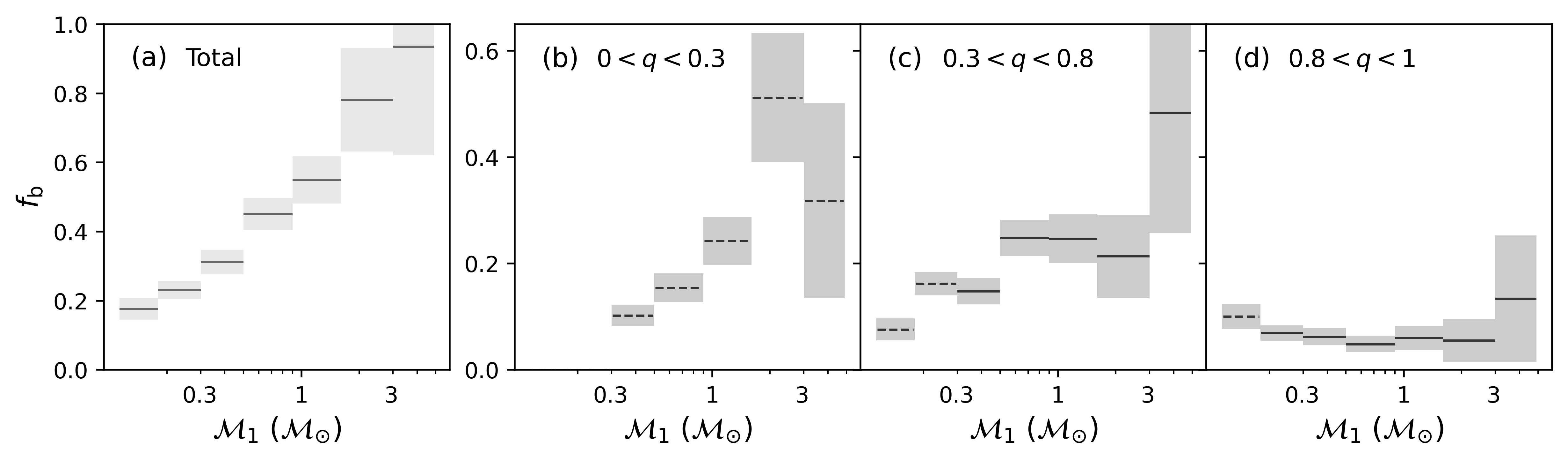}
    \caption{Binary fraction ($f_{\rm b}$) as a function of primary mass for different mass ratios in the Pleiades. Panel (a) shows the total binary fraction for $q>q_{\rm{lim}}$. Panels (b-d) represent the binary fractions for $q=0-0.3,~0.3-0.8,~0.8-1$, respectively. Dashed lines indicate incomplete binary fractions. The shaded regions represent $f_{\rm b}$ errors.}
    \label{fig:fb_m}
\end{figure*}

These comparisons demonstrate that the binary fraction in the Pleiades cluster varies significantly with the stellar mass and mass ratio ranges. Our work can reproduce or explain almost all previous results. This suggests that detection in the complete range of mass and mass ratios is extremely important in the analysis of binaries in clusters.

\subsection{Binary Fraction as a Function of Primary Mass}\label{sec:fb_m}

It is well established that $f_{\rm b}$ correlates with primary mass \citep{2020ApJ...901...49L,2023ASPC..534..275O}, indicating that stars with higher $\mathcal{M}_1$ are more likely to form and sustain binary systems. To explore this relation in the Pleiades, we divided the primary mass into seven segments: [0.11, 0.18, 0.3, 0.5, 0.9, 1.6, 3.0, 4.9] $\mathcal{M}_{\odot}$, roughly following a log-uniform dividing on $\mathcal{M}_1$.

The third column of Table~\ref{tab:fb_1} and panel (a) of Figure~\ref{fig:fb_m} show the variation of the total binary fraction ($f_{\rm b}$ for $q>q_{\rm lim}$) with primary mass. The $f_{\rm b}$ exhibits a monotonic increase with $\mathcal{M}_1$, ranging from 0.18 in the lowest mass bin to 0.94 within the highest mass bin. This trend confirms the previous conclusions that more massive stars are more likely to possess a companion. Meanwhile, our ability to detect very low mass-ratio companions of massive stars enhances this relationship. For example, we are able to identify binaries down to $q<0.03$, when $\mathcal{M}_1>3.0\mathcal{M}_{\odot}$. 

Furthermore, we divided the mass ratio into three ranges: $q=0-0.3,~0.3-0.8,$ and $0.8-1.0$ for the low, medium, and high mass ratios, and calculated $f_{\rm b}$ in these $q$ ranges with different primary masses. They are also listed in corresponding columns of Table~\ref{tab:fb_1}. For low and medium mass ratios, as shown in panels (b) and (c) of Figure~\ref{fig:fb_m}, the positive correlation between $\mathcal{M}_1$ and $f_{\rm b}$ remains. While for the high mass ratio, as shown in panel (d), $f_{\rm b}$ exhibits minimal variation with mass and lacks a significant trend. This behavior for high mass-ratio binaries is consistent with the result reported by \cite{2023AA...672A..29C}, who found no significant correlation between $f_{\rm b}$ and primary mass of $q>0.6$ binaries across 72 OCs, including the Pleiades.

In general, the variation of $f_{\rm b}$ with primary mass can be attributed to dynamical effects within the cluster \citep{2020ApJ...901...49L}.  Since the binary binding energy $E_{\rm{b}} \propto q\mathcal{M}_1^2$, binaries with lower primary masses and lower mass ratios are more easily disrupted during three-body encounters, leading to a decrease in $f_{\rm b}$ with decreasing $\mathcal{M}_1$ in the lower mass-ratio ranges. 

Alternatively, we propose another explanation for the $\mathcal{M}_1-f_{\rm b}$ relation of photometric binaries. As shown in Figure~\ref{fig:result}, the $q_{\rm lim}$ value is larger for lower-mass stars, meaning that many companions are too faint to be detected photometrically. In contrast, for higher-mass stars, a broader range of detectable mass ratios allows for the identification of more binaries. As observed in the Pleiades, many low mass-ratio binaries have a massive primary star (panel (b) of Figure~\ref{fig:fb_m}).

In summary, the variation in $f_{\rm b}$ with primary mass is influenced by both observational and dynamical effects. In the low mass-ratio range, the increase in $f_{\rm b}$ with primary mass is mainly driven by the observational limit of $q_{\rm lim}$. While in the medium mass-ratio range of $q = 0.3 - 0.8$, where the mass range of $\mathcal{M}_1 = 0.3 - 4.9 \mathcal{M}_{\odot}$ is unaffected by $q_{\rm lim}$, the rise in $f_{\rm b}$ with primary mass should be mainly influenced by dynamical effects.

\subsection{Binary Mass Ratio Distribution}\label{sec:qd}

The mass ratio distribution function of binaries, $n_{\rm b}(q)$, is a crucial statistic in star clusters. Similar to the stellar mass function, $n_{\rm b}(q)$ offers valuable insights into both the mechanism of star formation and the dynamical processes within a cluster.
\begin{figure}[h]
    \centering
	\includegraphics[width=0.94\columnwidth]{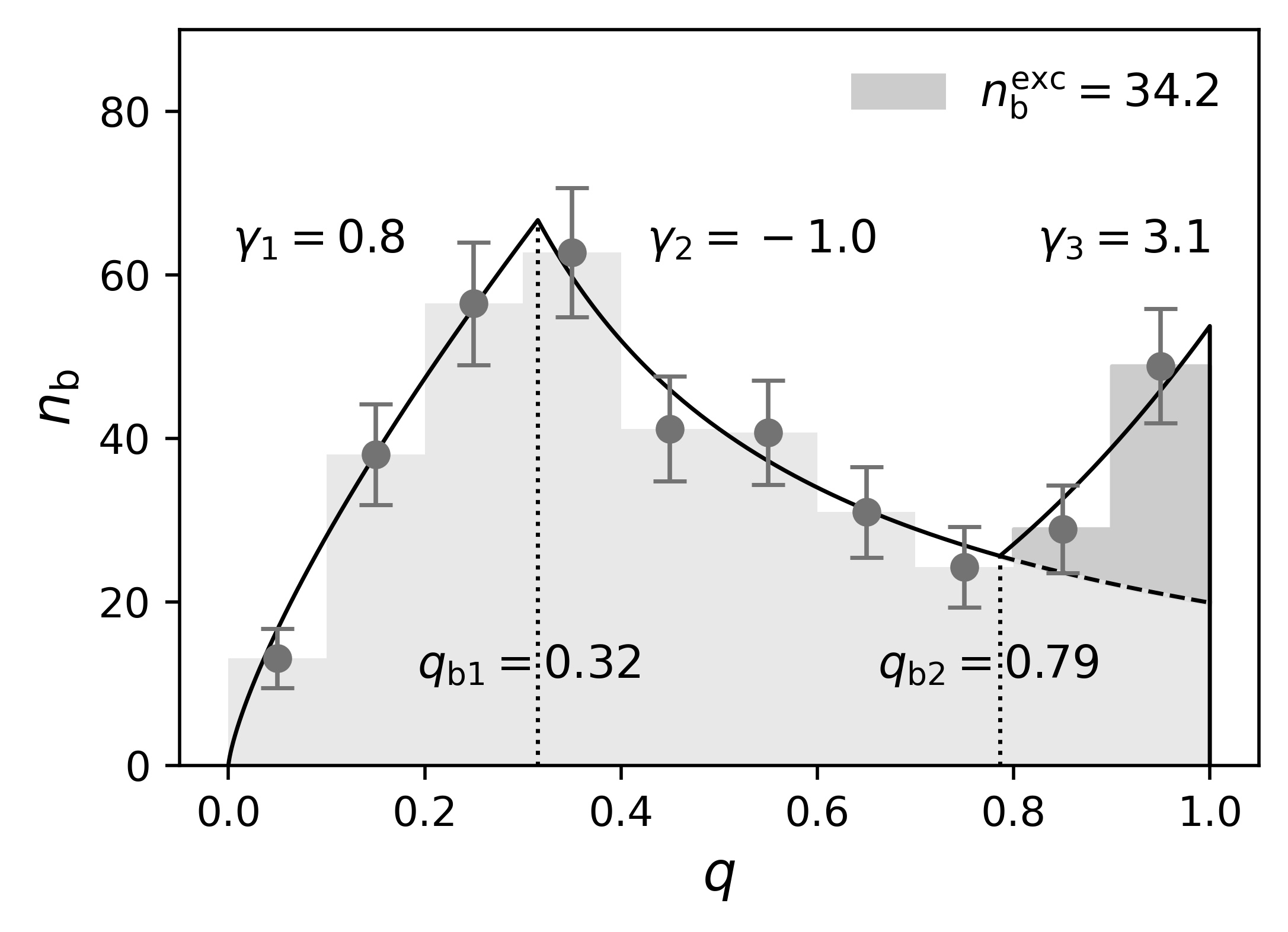}
    \caption{Mass ratio distribution and best-fit model for the Pleiades. The gray histogram and error bars show the mass ratio distribution. The black line represents the best-fit model. The model is divided into three segments. Each segment follows a power law. The dark gray area represents the number of exceeding high mass-ratio binaries.}
    \label{fig:qd_all}
\end{figure}

\begin{figure*}
    \centering
	\includegraphics[width=\textwidth]{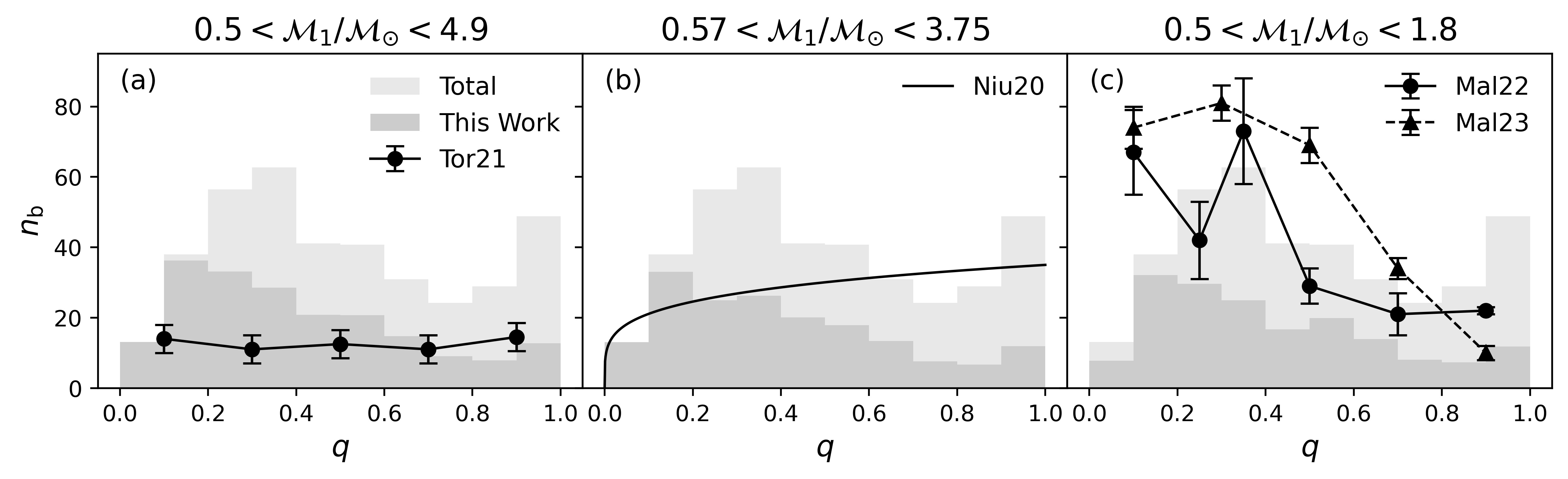}
    \caption{Comparison of mass ratio distribution from this work with other works. The light gray histograms represent the total mass ratio distribution across all mass ranges, whereas the dark gray ones represent the distributions within specific mass ranges for comparison. The solid line in panel (a) corresponds to the mass ratio distribution from \cite{2021ApJ...921..117T}. The solid line in panel (b) reflects the distribution from \cite{2020ApJ...903...93N}, which follows a power-law distribution with $\gamma = 0.22$. The solid and dashed lines in panel (c) denote the mass ratio distributions reported in \cite{2022AJ....163..113M} and \cite{2023AJ....165...45M}, respectively.}
    \label{fig:qd_compare}
\end{figure*}

Our measurement of the secondary mass reaches the lower limit of the PARSEC model at 0.09$\mathcal{M}_{\odot}$, enabling us to obtain a board mass-ratio distribution above $q_{\rm lim}$. The gray histogram in Figure~\ref{fig:qd_all} shows the $n_{\rm b}(q)$ for all photometric binaries in our sample, with the primary mass ranging from 0.11 to 4.9$\mathcal{M}_{\odot}$. The profile of $n_{\rm b}(q)$ can be divided into three segments. It increases in the low mass-ratio range ($q\lesssim 0.3$), subsequently decreases in the medium mass-ratio range ($0.3 \lesssim q \lesssim 0.8$) and then increases again in the high mass-ratio range ($q\gtrsim 0.8$). We modeled this distribution using a three-segment power-law profile, uniformly expressed as $n_{\rm b}(q)\propto q^\gamma$. The power-law exponents for three segments, ranging from low to high mass ratios, are as follows: $\gamma_1=0.8 \pm 0.2$, $\gamma_2=-1.0 \pm 0.3$, and $\gamma_3=3.1 \pm 1.0$. This modeling identifies two breakpoints of mass ratio: $q_{\rm b1} = 0.32 \pm 0.04$, $q_{\rm b2} = 0.79 \pm 0.05$. This three-segment profile of the Pleiades resembles the mass ratio distribution model proposed by \cite{2017ApJS..230...15M} (See Figure~2 of their paper). The difference is that their model shows a significant excess of twin binaries, while the Pleiades appears a progressive excess of high mass-ratio binaries. If we extrapolate $\gamma_2$ to the high mass-ratio range, then the number of exceeding binaries is estimated to be $n_{\rm b}^{\rm exc} \approx 34.2\pm 5.8$.

Several studies have investigated the mass ratio distribution of the Pleiades, mostly focused on binaries with $\mathcal{M}_1 \gtrsim 0.5 \mathcal{M}_{\odot}$, with varying upper mass limits. Figure~\ref{fig:qd_compare} compares our results with others. The light gray histograms represent the mass ratio distribution for all binaries as the same as in Figure~\ref{fig:qd_all},  while the dark gray ones are constrained to the same mass ranges of the compared works. The solid line in panel (a) shows a nearly flat mass ratio distribution obtained by \cite{2021ApJ...921..117T} from the $\Delta RV$ method. Compared to our result, there is a significant reduction in the number of low mass-ratio binaries, likely due to the limited sensitivity of the spectroscopic in this mass-ratio range. The solid line in panel (b) shows the result from \cite{2020ApJ...903...93N}, who applied a mixture-model method on the CMD with Gaia photometry. They employed a power-law profile but had an index of 0.22, indicating a roughly flat distribution. Their result also shows fewer low mass-ratio binaries than ours. This difference may arise from two factors. One is the use of optical data alone, which may inadequately distinguish low mass-ratio binaries from single stars. Another is the absence of isochrone correction, potentially introducing bias in the measurement of binary mass ratios. The solid and dashed lines in panel (c) represent results from \cite{2022AJ....163..113M} and \cite{2023AJ....165...45M}, who used a pseudo-color diagram combining optical to infrared photometric data. Their mass ratio distribution reveals a higher proportion of low mass-ratio binaries compared to high mass-ratio ones. This phenomenon is consistent with our result. However, they have identified more binaries than ours, as we have already mentioned and discussed in Figure~\ref{fig:fb_compare} of Section~\ref{sec:fb}.

\begin{figure*}
    \centering
	\includegraphics[width=\textwidth]{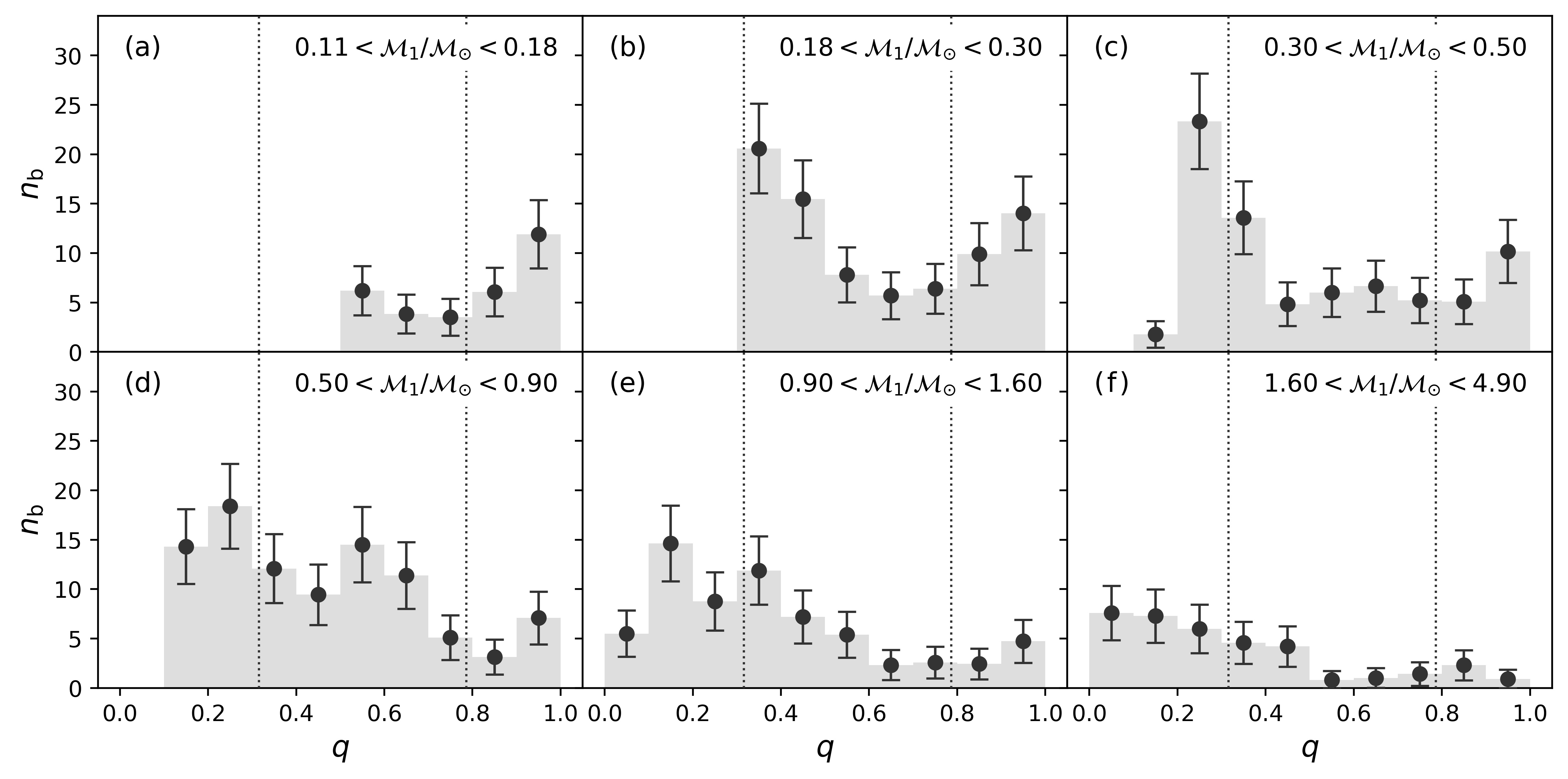}
    \caption{Mass ratio distributions in different mass ranges. Panels (a-f) show distributions in different mass ranges: $\mathcal{M}_1=0.11-0.18,~0.18-0.3,~0.3-0.5,~0.5-0.9,~0.9-1.6,~1.6-4.9 \mathcal{M}_{\odot}$. Within each panel, the dotted lines indicate the peak and valley of the total mass ratio distribution.}
    \label{fig:qd_M1}
\end{figure*}

In conclusion, the mass-ratio distribution of photometric binary stars in the Pleiades shows significant complexity. Therefore, using a single power law is inadequate to accurately characterize this distribution.

\subsection{Mass Ratio Distribution Varying with Primary Mass}\label{sec:qd_m}

By analyzing subsamples with various primary masses, we aim to conduct a more detailed investigation into the mass ratio distribution of binaries and its underlying causes. The mass binning follows the scheme presented in Section~\ref{sec:fb_m}, except that the two bins of $\mathcal{M}_1 > 1.6\mathcal{M}_{\odot}$ are combined. The segmentation points of $\mathcal{M}_1$ are [0.11, 0.18, 0.3, 0.5, 0.9, 1.6, 4.9] $\mathcal{M}_{\odot}$.

Figure~\ref{fig:qd_M1} presents the mass ratio distribution within different primary mass ranges. Several key features are observed. First, $n_{\rm b}(q)$ exhibits significant variations across different mass ranges, and most of them do not follow a monotonic trend. Second, as the primary mass increases, the position of $q_{\rm b1}$ tends to shift towards lower mass ratios. This strongly suggests that, at $q<0.8$, the mass ratio of real binary systems, whether or not having an illuminated companion, might be monotonically decreasing. The observed break of photometric binaries ($q_{b1}$) attributed to the selection effects caused by the detection limit, $q_{\rm lim}(\mathcal{M}_1)$. Third, at lower $\mathcal{M}_1$, the number of binaries with high-$q$ exceeds that of binaries with medium-$q$, suggesting that high-$q$ binaries are predominantly composed of low-mass stars with $\mathcal{M}_1 < 0.5\mathcal{M}_{\odot}$. This pattern is also evident in the two-dimensional PDF shown in panel (b) of Figure~\ref{fig:result}. Specifically, the probability density decreases as $q$ moves away from $q_{\rm lim}$, reflecting the monotonic decrease in $n_{\rm b}(q)$ for binaries with low and medium mass ratios. Additionally, a prominent high-density region is observed in the upper left of this panel, corresponding to an excess of low-$\mathcal{M}_1$ and high-$q$ binaries.

The mass-ratio distribution of binaries is shaped by various factors, and the complex profile observed in the Pleiades cluster reflects the combined impact. Besides the selection effect imposed by the $q_{\rm lim}(\mathcal{M}_1)$, the main physical reasons lie in the star formation mechanisms and the dynamical evolution processes. The value of $\gamma_2=-1.0$ is notably related to the stellar mass function. However, this exponent index implies the absence of low mass-ratio binaries compared to random pairings from the initial mass function ($\alpha_{\rm MF}\sim -2.35$). It may be caused by the dynamics that result in low-$q$ binaries more susceptible to disruptions in three-body encounters. Additionally, the excess of high-$q$ binaries can be explained by two factors. First, during the formation of primordial binaries, the accretion rate of the secondary star is typically higher than that of the primary star in the competitive accretion mechanism within binary discs \citep{2000MNRAS.314...33B,2014ApJ...783..134F,2019ApJ...871...36M}. This leads to the secondary star having a mass comparable to the primary star's. Second, during subsequent dynamical interaction in the cluster, the lower-mass companions are more likely to be replaced by a higher-mass single star, leading to a higher mass-ratio binary. This effect is particularly pronounced in binaries with low-mass primaries due to their lower binding energies and the abundance of single stars with comparable masses available for exchange.

\section{Conclusion}\label{sec:conclusion}

We present an optimized multiband magnitude fitting approach for determining the stellar mass of main-sequence single stars or binary components in a cluster. Employing photometric data from Gaia and 2MASS, we have measured the stellar mass of members in the Pleiades open cluster and performed a comprehensive analysis of its main-sequence binaries, including the binary fraction, mass-ratio distribution, as well as their relationships with the primary mass.

The improvements and advantages of our fitting method are as follows:
\begin{enumerate}

\item By utilizing the concentrated distribution of single stars along the main sequence, we have adjusted the theoretical model to match the photometric data, resulting in an empirical photometric model. This adjustment significantly reduces biases in binary mass-ratio measurements.

\item The combined use of optical and near-infrared photometric data covers the primary emission bands of stars with a wide mass range. This significantly enhances the precision of mass fitting, particularly for low-mass stars, which is crucial for the identification and accurate quantification of low mass-ratio binaries.

\item The Bayesian framework allows us to rigorously quantify stellar masses and mass ratios using probabilistic methods, enabling our detection down to the theoretical lower mass limit. By stacking the probability density functions (PDFs) of individual members in a cluster, we can extend the Bayesian approach to the analysis of binary properties across the entire sample.

\end{enumerate}

By applying this method to the Pleiades cluster, we detected and measured all photometric main-sequence binaries with mass ratios above a well-defined limit, $q_{\rm lim}(\mathcal{M}_1)$. The detailed results of these binary properties are listed below:

\begin{enumerate}

\item The binary fraction of main-sequence members in the Pleiades cluster is $0.34 \pm 0.02$. 

\item Observations show that the binary fraction increases with increasing primary mass. This trend is particularly evident in binaries with low and intermediate mass ratios.

\item The complete mass ratio distribution follows a three-segment power-law profile, with power-law exponents ($\gamma$) to be $0.8\pm 0.2$, $-1.0\pm 0.3$, and $3.1 \pm 1.0$ for lower, median and higher mass-ratio ranges that separated by two break points at $q\sim$ 0.3 and 0.8.  This pattern indicates a generally decreasing trend, with a significant deficiency of binaries in the low mass-ratio range and a notable excess in the high mass-ratio range.

\item By analyzing subsamples with different primary masses, we find that the proportion of high mass-ratio binaries decreases as the primary mass increases. This indicates that these photometric high mass-ratio binaries are mainly contributed by binaries with lower primary masses.

\item As a by-product, we found that, for the current sample from Gaia DR3, almost all single stars have RUWE$<$1.4, while only about $1/3$ of binaries are larger than this value.

\end{enumerate}

These results provide a comprehensive overview of the statistical properties of photometric binaries in the Pleiades cluster, revealing complex patterns in the distribution of binary fraction and mass ratio. The complexity can be attributed to multiple factors, including detection biases for low-mass binaries, binary formation mechanisms in the star-forming stage, and the subsequent dynamical interactions within the cluster.

\section*{Acknowledgments}
We sincerely thank the anonymous referee for valuable comments and suggestions. This work was supported by the National Natural Science Foundation of China (NSFC) under grants 12273091 and U2031139; the National Key R\&D Program of China No.~2019YFA0405501; the science research grants from the China Manned Space Project with No.~CMS-CSST-2021-A08; the Science and Technology Commission of Shanghai Municipality (Grant No.~22dz1202400); the Program of Shanghai Academic/Technology Research Leader. This work has made use of data from the European Space Agency (ESA) mission Gaia (https://www.cosmos.esa.int/gaia), processed by the Gaia Data Processing and Analysis Consortium (DPAC; https://www.cosmos.esa.int/web/gaia/dpac/consortium). Funding for the DPAC has been provided by national institutions, in particular the institutions participating in the Gaia Multilateral Agreement. This work made use of data products from the Two Micron All Sky Survey, which is a joint project of the University of Massachusetts and the Infrared Processing and Analysis Center/California Institute of Technology, funded by the National Aeronautics and Space Administration and the National Science Foundation.

\software{PARSEC \citep{2012MNRAS.427..127B},
astropy \citep{2013A&A...558A..33A,2018AJ....156..123A},  
           Nautilus \citep{nautilus}}







\bibliography{reference}{}
\bibliographystyle{aasjournal}



\appendix 

\section{Bayesian Inference of Parallax for Nearby Cluster Members}
\label{apen:BayesPlx}

For nearby star clusters, both the scale of the line-of-sight distribution and the parallax measurement uncertainties of individual members are non-negligible. Simply adopting the average parallax of the cluster would ignore the distance variations among individual members. Alternatively, if we directly use the observational parallax of each member, due to the typically large uncertainties, additional scatter would be introduced into the line-of-sight distribution of the cluster. For instance, when using Gaia parallaxes directly, one can often observe an apparent elongation of the nearby cluster along the line of sight.

Based on the Bayesian Inference, we introduce a method to calculate the more reliable parallax for an object if it is confirmed to be a cluster member. The posterior probability distribution function (PDF) of the modified parallax of this object within the cluster can be expressed as:
\begin{equation}
    \mathcal{P}(\varpi | \varpi_{\rm o}) \propto \mathcal{L}(\varpi_{\rm o} | \varpi) \cdot \pi(\varpi) .
\end{equation}
The $\pi(\varpi)$ represents the real $\varpi$ distribution of cluster members. Considering that the scale of a star cluster is generally much smaller than its distance, we can reasonably assume that the $\pi(\varpi)$ follows a normal distribution,  
\begin{equation}
    \pi(\varpi) \propto \mathcal{N}(\varpi_{\rm c}, \sigma_{\varpi_{\rm c}}), 
\end{equation}
where $\varpi_{\rm C}$ is the mean parallax of the cluster, and $\sigma_{\varpi_{\rm C}}$ represents the intrinsic scatter of the parallax distribution of cluster members. This scatter can be characterized by the intrinsic dispersion of observed parallaxes or, under the assumption of spherical symmetry, it can be estimated using the projected size of the cluster. The $\mathcal{L}(\varpi_{\rm o} | \varpi)$ is the likely probability distribution according to the parallax observation of the concerned member, which also follows a normal distribution,
\begin{equation}
    \mathcal{L}(\varpi_{\rm o} | \varpi)  \propto \mathcal{N}(\varpi_{\rm o}, \sigma_{\varpi_{\rm o}}), 
\end{equation}
with $\varpi_{\rm o}$ and $\sigma_{\varpi_{\rm o}}$ to be the observational parallax and the uncertainty of this object.

It can be rigorously proved that the $\mathcal{P}(\varpi | \varpi_{\rm o}) $ also obeys a normal distribution, 
\begin{equation}
    \mathcal{P}(\varpi | \varpi_{\rm o})  \propto \mathcal{N}(\varpi_{\rm m}, \sigma_{\varpi_{\rm m}}), 
\end{equation}
with the modified parallax ($\varpi_{\rm m}$) and the uncertainty ($\sigma_{\varpi_{\rm m}}$) are derived as:
\begin{equation}
    \varpi_{\rm m}=(\varpi_{\rm c} \sigma_{\varpi_{\rm c}}^{-2} + \varpi_{\rm o}\sigma_{\varpi_{\rm o}}^{-2})/(\sigma_{\varpi_{\rm c}}^{-2} + \sigma_{\varpi_{\rm o}}^{-2})       , 
\end{equation}
and 
\begin{equation}
    \sigma_{\varpi_{\rm m}} = (\sigma_{\varpi_{\rm c}}^{-2} + \sigma_{\varpi_{\rm o}}^{-2} )^{-1/2} .
\end{equation}

For the Pleiades cluster of the present work, the $\varpi_{\rm c}=7.41$ mas is the average value of our kinematic member sample. The $\sigma_{\varpi_{\rm c}}$ is adopted as the intrinsic dispersion fitted from this sample, equals 0.17 mas. This line-of-sight scale is roughly equivalent to the projected scale of the Pleiades.

\end{document}